# The OSIRIS-REx Thermal Emission Spectrometer (OTES) Instrument

P.R. Christensen[1], V.E. Hamilton[2], G.L. Mehall[1], D. Pelham[1], W. O'Donnell[1], S. Anwar[1], H. Bowles[1], S. Chase[1], J. Fahlgren[1], Z. Farkas[1], T. Fisher[1], O. James[3], I. Kubik[1], I. Lazbin[4], M. Miner[1], M. Rassas[1], L. Schulze[1], K. Shamordola[1], T. Tourville[1], G. West[4], R. Woodward[1], D. Lauretta[5]

[1]*School of Earth and Space Exploration, Arizona State University, Tempe, AZ, USA*
Phone:  480-965-7105
(phil.christensen@asu.edu)

[2]*Southwest Research Institute, Boulder, CO, USA*

[3]*Goddard Space Flight Center, Greenbelt, MD, USA*

[4]*Moog Broad Reach, Tempe, AZ; currently at AZ Space Technologies, Gilbert, AZ, USA*

[5]*University of Arizona, Tucson, AZ, USA*

**Abstract** The OSIRIS-REx Thermal Emission Spectrometer (OTES) will provide remote measurements of mineralogy and thermophysical properties of Bennu to map its surface, help select the OSIRIS-REx sampling site, and investigate the Yarkovsky effect. OTES is a Fourier Transform spectrometer covering the spectral range 5.71–100 μm (1750–100 cm$^{-1}$) with a spectral sample interval of 8.66 cm$^{-1}$ and a 6.5-mrad field of view. The OTES telescope is a 15.2-cm diameter Cassegrain telescope that feeds a flat-plate Michelson moving mirror mounted on a linear voice-coil motor assembly. A single uncooled deuterated L-alanine doped triglycine sulfate (DLATGS) pyroelectric detector is used to sample the interferogram every two seconds. Redundant ~0.855 μm laser diodes are used in a metrology interferometer to provide precise moving mirror control and IR sampling at 772 Hz. The beamsplitter is a 38-mm diameter, 1-mm thick chemical vapor deposited diamond with an antireflection microstructure to minimize surface reflection. An internal calibration cone blackbody target provides radiometric calibration. The radiometric precision in a single spectrum is ≤2.2 × 10$^{-8}$ W cm$^{-2}$ sr$^{-1}$ /cm$^{-1}$ between 300 and 1350 cm$^{-1}$. The absolute integrated radiance error is <1% for scene temperatures ranging from 150 to 380 K. The overall OTES envelope size is 37.5 × 28.9 × 52.2 cm, and the mass is 6.27 kg. The power consumption is 10.8 W average. The OTES was developed by Arizona State University with Moog Broad Reach developing the electronics. OTES was integrated, tested, and radiometrically calibrated on the Arizona State University campus in Tempe, AZ.

## Abbreviations

| | |
|---|---|
| ADC | analog to digital converter |
| ARM | antireflection microstructure |
| ASU | Arizona State University |
| BCU | Bench Checkout Units |
| CDR | Critical Design Review |
| CPT | Comprehensive Performance Test |
| CVD | chemical vapor deposited |
| DFT | discrete Fourier transform |
| DLATGS | deuterated L-alanine doped triglycine sulfate |
| EM | Engineering Model |
| FET | Field Effect Transistor |
| FPGA | Field Programmable Gate Array |
| FPU | Field Programmable Unit |
| FWHM | full-width, half-maximum |
| GSE | Ground Support Equipment |
| IFOV | instantaneous field of view |
| IFT | Instrument Functional Test |
| IMU | inertial measure unit |
| IRF | instrument response function |
| LMSS | Lockheed Martin Space Systems |
| MER | Mars Exploration Rover |
| MGS | Mars Global Surveyor |
| Mini-TES | Miniature TES |
| MLI | Multi-Layer Insulation |
| MO | Mars Observer |
| NEΔε | Noise Equivalent Delta Emissivity |
| NESR | noise equivalent spectral radiance |
| OTES | OSIRIS-REx Thermal Emission Spectrometer |
| PDR | Preliminary Design Review |
| S/C | spacecraft |
| SNR | signal-to-noise ratio |
| SPOC | Science Processing and Operations Center |
| SRR | System Requirement Review |
| TES | Thermal Emission Spectrometer |
| THEMIS | Thermal Emission Imaging System |
| ZPD | zero path difference |

# 1 Introduction

The OSIRIS-REx Thermal Emission Spectrometer (OTES) instrument will aid in the characterization and sample site selection of the asteroid Bennu (101955 Bennu) for the OSIRIS-REx sample return mission (Lauretta et al. 2017) through the determination of Bennu's mineralogy and thermophysical properties. The specific science objectives of the OTES investigation are to: (1) document the sample site spectral properties and mineralogy/geochemistry; (2) globally map the distribution of materials on Bennu's surface; (3) determine regolith physical properties (e.g., grain size and subsurface layering) using diurnal temperature measurements; (4) measure thermal emission to determine properties that contribute to the Yarkovsky effect; and (5) search for and characterize volatile outgassing, satellites, and space weathering.

These objectives are addressed using thermal infrared spectral observations between 1750 and 100 cm$^{-1}$ (5.7 to 100 µm). The OTES spectral range, spectral sampling, radiometric resolution, and spatial resolution provide high-quality, mid-infrared spectra at sufficient ground resolution to investigate the mineralogy and physical properties of small-scale surface features on Bennu and to aid in the selection of the OSIRIS-REx sampling site. Figure 1 gives representative meteorite and mineral spectra of materials that may be present on Bennu. From this figure it is apparent that the thermal IR spectral range contains a wealth of information, which is unique to each mineral, that will be used to identify and map minerals across Bennu's surface.

In this paper we present a description of the OTES as-built instrument, calibration methods and knowledge, and system performance. We also describe the basic instrument operational strategy, data processing methodology, and the plans for archiving the data through the Planetary Data System.

[Insert Figure 1]

## 2 Instrument Description

### 2.1 Instrument Overview

The OTES instrument follows a progression of infrared spectrometers and multi-spectral imagers developed by Arizona State University (ASU) for use on an array of planetary missions, including the Mars Observer (MO) and Mars Global Surveyor (MGS) Thermal Emission Spectrometer (TES) instruments (Christensen et al. 1992, 2001), the two Mars Exploration Rover (MER) Miniature TES (Mini-TES) instruments (Christensen et al. 2003), and the Mars Odyssey Thermal Emission Imaging System (THEMIS) instrument (Christensen et al. 2004a). These instruments were inspired by the pioneering work of R.A. Hanel and colleagues in a series of missions beginning with Nimbus III in 1969 (Hanel et al. 1970) and extending through the Mariner 9 (Hanel et al. 1970) and Voyager planetary missions (Hanel et al. 1980). Table 1 provides a comparison of the ASU instruments, and shows the progression in their performance.

[Insert Table 1]

OTES is a Fourier-transform interferometer that collects hyperspectral thermal infrared data over the spectral range from 1750 to 100 cm$^{-1}$ (5.7 to >100 µm) with a spectral sampling of 8.66 cm$^{-1}$ and a 6.5-mrad full-width, half-maximum (FWHM) instantaneous field of view (IFOV) in a single detector. The OTES wavelength range and spectral resolution are sufficient to resolve the key vibrational absorption features of silicate, carbonate, sulfate, phosphate, oxide, and hydroxide minerals. The design of the OTES is intentionally conservative and utilizes the identical interferometer design used on previous ASU infrared spectrometers. The electronics and interferometer control have been redesigned from the most recent instrument in this series, the MER Mini-TES instrument (Christensen et al. 2003), in order to utilize updated electronic components and incorporate a digital servomechanism control.

The primary measurement requirement is to obtain spectra with sufficient resolution and radiometric precision to allow materials of different mineralogy to be characterized and distinguished and to allow determination of relative abundances in mineral mixtures. In addition, it is necessary to have sufficient radiometric accuracy to allow the thermal inertia and thermal radiance to be determined with sufficient accuracy to characterize potential sample sites and the Yarkovsky effect (e.g., Bottke et al. 2006). During the approach, reconnaissance, and detailed survey phases of the OSIRIS-REx mission (Lauretta et al. 2017) OTES will provide full-disk integrated spectral data, global spectral maps at 40-m resolution, and local spectral information of the sample site at ~4-m resolution, respectively. These observations will provide the capability to determine and map the mineral composition and thermal inertia of the asteroid surface rocks and soils, which will be important for identifying the optimum region for sampling from both a scientific and sampling safety perspective.

## 2.2 OTES Measurement Requirements

### 2.2.1 Composition

The compositional requirement is for OTES to determine mineral abundances in mixtures to ±10% accuracy. Typical absorption band emissivity minimums in mixtures of medium-grained (~100 µm) particulates and for moderate absorption bands are 0.95, corresponding to a band depth of 0.05 emissivity. Extensive experience with lab and spacecraft data has shown that in order to identify components OTES requires a radiometric precision of 1/16 of the band depth (or a "science to noise ratio" of 16) (e.g., Ramsey and Christensen 1998; Hamilton and Christensen 2000; Christensen et al. 2001, 2008; Bandfield et al. 2004; Rogers and Christensen 2007). This requirement produces an emissivity precision requirement, stated as the Noise Equivalent Delta Emissivity (NE$\Delta\varepsilon$), equal to $0.05 \times 1/16 = 0.003125$. The NE$\Delta\varepsilon$ is equal to the inverse of the signal-to-noise ratio (SNR), giving an SNR requirement of 320. Laboratory and spacecraft experience has also shown that the absolute emissivity requirement is ~10% of the emissivity, primarily to minimize spectral slopes, giving an absolute emissivity requirement of $0.9 \times 0.1 = 0.09$. This absolute requirement derived for compositional measurements is far less stringent than that derived from the Yarkovsky effect.

The signal-to-noise ratio varies as a function of scene temperature and wavenumber. Therefore, the 1-sigma radiometric precision, or noise equivalent spectral radiance (NESR) that is required to produce a given SNR also varies with scene temperature and wavenumber. For ease of verification during test and calibration, the formal OTES SNR requirement was set to be 320 at a reference temperature of 325 K between 300 and 1350 cm$^{-1}$, which corresponds to an NESR requirement of ≤2.3 × 10$^{-8}$ W cm$^{-2}$ sr$^{-1}$ /cm$^{-1}$ over this spectral range.

### 2.2.2 Yarkovsky Effect

The OTES Yarkovsky effect requirement is an accuracy of 1.5% of the total emitted radiance from 6 to 50 μm at the time of instrument delivery. The primary sources of accuracy errors in the OTES radiance are the error in the absolute knowledge of the internal calibration blackbody temperature and emissivity, the reflectivity of the calibration flag mirror, and the temperature and emissivity of the fore optics [see Section 3.2]. This requirement has been converted to absolute integrated radiance requirement as a function of scene temperature in Table 2.
[Insert Table 2]

### 2.2.3 Thermal Inertia

The OSIRIS-REx thermal inertia requirement is to determine thermal inertia to ±10 J m$^{-2}$ K$^{-1}$ s$^{-½}$ accuracy. This requirement translates to an OTES requirement to determine the absolute surface temperature to ±2 K for typical night (220 K) and day (350 K) temperatures, and ±5 K for the extreme polar (70 K) temperatures. The thermal inertia error sources are the same as those for the Yarkovsky effect.

## 2.3 OTES Design

OTES builds heavily upon the previous TES and Mini-TES instruments (Christensen et al. 1992, 2001, 2003). The interferometer moving mirror mechanism is identical to these previous instruments, the telescope size and aperture are similar to TES, and the detector is similar to both instruments. The primary new developments were the use of diamond, in place of CsI or KBr, as the IR beamsplitter substrate material, and a digital servomechanism for the Michelson moving mirror control. OTES was designed to be modular for ease in application to future missions, and consists of a telescope, interferometer assembly, control electronics, internal calibration blackbody, and structure (Fig. 2). The functional block diagram is shown in Figure 3, and the instrument properties are summarized in Table 3.
[Insert Table 3]
[Insert Figure 2]

[Insert Figure 3]

*2.3.1 Opto/Mechanical*

The OTES optical system uses a compact Cassegrain telescope design with a 15.2-cm diameter f/3.91 Ritchey-Chretien telescope that feeds a flat-plate Michelson moving mirror mounted on a voice-coil motor assembly (Fig. 4). An off-axis parabolic mirror converts the telescope output beam to an optical beam with an afocal ratio of eight that passes through the interferometer. This radiance passes through a diamond beamsplitter installed in a radial three-point mount. Half of the radiance is passed to a fixed mirror, the other half to the moving mirror, which travels ±0.289 mm to achieve 8.66 cm$^{-1}$ spectral sampling. These two beams recombine at the beamsplitter, producing the interference that is used to determine the spectral distribution of the scene radiance, and travel to a parabolic mirror that images the beam onto a single on-axis detector element with a chemical vapor deposited (CVD) diamond lens. Two redundant monochromatic ULM VCSEL laser diodes with a nominal wavelength of 0.854 ±.002 µm feed a fringe counting metrology interferometer that uses the same moving and fixed mirrors and beamsplitter as used in the IR signal chain. This metrology interferometer is used to precisely control the velocity, track the position of the moving mirror, and trigger the sampling of the IR signal. The laser diodes provide four times oversampling of the shortest scene bandpass wavelength. The precise wavelength of these lasers was determined using observations of known spectral samples (Section 6.4). All mirror surfaces are diamond-turned and gold-coated, with reflectances of >0.99 across the OTES spectral range. OTES is aluminum for light weight and strength while meeting the launch load and pyro shock requirements. The use of baffles around the telescope housing and secondary mirror, and the use of diffuse black Z306 paint around the optics and within the cavity, minimize stray light effects.

[Insert Figure 4]

OTES utilizes a 38-mm-diameter, 1-mm-thick CVD diamond beamsplitter substrate. This diamond was fabricated by Diamond Materials and is installed in a radial three-point mount that provides precise alignment over the 10 °C to 40 °C operational range, maintains mechanical integrity over the –25 °C to +55 °C nonoperational proto-flight survival range, and accommodates the low thermal expansion coefficient and high heat conductivity of diamond. CVD growth and polishing advancements in the last 10 years provided the optical performance to meet the OTES requirements. An antireflection microstructure (ARM) developed by TelAztec provides surface reflectances of >78% over the spectral range from 7 to >25 µm. The low dispersion properties of diamond allow a single substrate to be used without the need for a compensator component, which would have increased the number of surface reflections and reduced the instrument throughput. The only significant disadvantage of diamond is the strong absorption feature between 1800 and 2700 cm$^{-1}$ (5.6 and 3.7 µm) (Coe and Sussmann 2000), which limits the effective spectral range of OTES to <1800 cm$^{-1}$ (>5.6 µm). The IR beamsplitting function is provided by Ge beam-dividing coating. A parabolic mirror reimages the optical pupil onto an on-axis thermal IR detector that has a small (4.4 mm) CVD diamond lens also fabricated by Diamond Materials.

### 2.3.2 Detector

OTES uses an uncooled deuterated L-alanine doped triglycine sulfate (DLATGS) pyroelectric detector fabricated by Selex-Galileo. The measured D*, including the diamond lens, is $1.2 \times 10^9$ cm•Hz$^{1/2}$/watt at 10 Hz, 22 °C. The detector has a bias voltage applied to the Field Effect Transistor (FET) to ensure it remains properly poled. Similar detectors were used on the MER Mini-TES instruments. Preamplification and front-end filtering are performed on the preamplifier circuit board to amplify the signal and to AC couple the detector output to block high-frequency oscillations.

### 2.3.3 Electronics

A ±12-volt regulator supplies power to the detector and preamplifier electronics. The spectrometer circuit board performs the bulk of the analog electronics processing. OTES command, control, and data flow tasks are controlled by logic in the command and control FPGA. The interface electronics parse out the instrument commands that control the OTES hardware functions. The flow of the digital interferometer data is controlled by additional logic in the command and control board Field Programmable Gate Array (FPGA). After each interferometer scan, the 16-bit interferogram data and 16-bit telemetry data are moved from the analog to digital converter (ADC) to the input memory buffer on a 16-bit parallel data bus. The 16-bit data are then serialized for transfer to the spacecraft (S/C) via RS-422 interface. The analog multiplexer, digital serializer, and data formatting logic are included in the command and control FPGA. The DC power converter accepts 24 to 34 volts unregulated input voltage and supplies +3.3, ±5 and ±15 volts regulated output voltage.

The OTES timing sequencing electronics are implemented in the command and control board FPGA. These electronics generate the timing waveforms necessary to control and synchronize instrument operation and provide the control and synchronization of the amplification, track/hold, multiplexing, and analog to digital conversion of the detector and telemetry signals. They also control and synchronize the interferometer servo electronics with the data acquisitions. The timing sequencing electronics include the fringe delay electronics that are used to correct the sampling error due to the phase delays between the fringe and IR analog channels. All clocks in the timing sequencer are generated from the master clock crystal oscillator that operates at a frequency of 20 MHz. The OTES interferometer servo electronics were completely redesigned from previous instruments to use a digital servo drive running in the FPGA. This digital servo receives scan timing clocks from the timing sequencer electronics and the fringe clock from the fringe detection electronics. The motor control logic uses these clocks to synchronize the mirror movement with the spectrometer data acquisitions. The moving mirror uses a direct drive BEI Kimco linear motor with tachometer feedback. The moving mirror tachometer signal is returned to the interferometer control electronics to allow active feedback control of the actuator.

The digital servo control loop is implemented using a custom processor core with a double precision Field Programmable Unit (FPU) that runs inside an FPGA mounted on the servo control board. The control loop uses digitized metrology fringe, tachometer, and end-of-travel optical switch as inputs; performs fringe data processing, feedback compensation, sample timing, and command/telemetry functions; and outputs the motor command to the D/A converter controlling the voice-coil driver circuit for the interferometer moving mirror.

High sample frequency of the fringe signal (20 kHz) enables the control loop to estimate the servo velocity at a higher frequency than would be achievable by only monitoring the metrology interferometer zero crossings. A rate estimation algorithm inside the servo controller uses the sinusoid model of the metrology fringe to derive a servo rate estimate at a 2-kHz rate. This model is also used to predict the appropriate times to sample the interferometer detector. The detector samples are based on estimated fringe signal zero crossings and are delayed to compensate for the phase delay of the Bessel filter in the detector electronics. An optical switch is used near the end of travel to measure the absolute position and control the servo motion so that it remains centered about the zero path difference (ZPD) position. The control algorithm includes adjustable optical switch thresholds so that the scan pattern can be centered without physically moving the optical switch assembly. The control circuit also includes a "gravity override" mode, which applies extra motor current to oppose the force of gravity so that the instrument can be tested in any orientation while on the ground.

There is a great deal of flexibility in the design and adjustment of the digital servo. Controller gains and limits, as well as some other configuration options, such as parameters controlling scan timing, can be changed via commands while the controller is executing out of SRAM. Also, because the entire algorithm is stored in nonvolatile memory, it can be modified without changes to hardware.

The flow of the digital interferometer data is controlled by additional logic in the command and control board FPGA. After each interferometer scan, the 16-bit interferogram data and 16-bit telemetry data are moved from the A/D to the input memory buffer on a 16-bit parallel data bus.

### 2.3.4 Thermal Design

OTES is a thermal instrument that requires thermal stability of several °C over tens of minutes to maintain calibration between periodic calibration observations. A temperature stability of <0.1 °C per minute is achieved by conductively isolating OTES from the spacecraft to prevent rapid thermal transients, while radiatively coupling OTES to the spacecraft to provide a long-term thermal sink. Replacement heaters and thermal blanketing are used when the instrument is off to offset the thermal loss through the OTES aperture and to maintain the instrument within its flight allowable temperature limits. The OTES will operate within specification over a temperature range of 10 to 40 °C, operate out of spec from –15 to +45 °C, and survive form –25 to +55 °C.

### 2.3.5 S/C Interface

The OTES spectrometer provides data to the spacecraft across two synchronous RS-422 serial lines at 57.6 Kbps during the 200-msec interferometer scan reversal period whenever OTES is powered. Instrument commanding, including operation of the internal cal mirror and the selection of redundant components, is provided across two synchronous RS-422 serial lines at 57.6 Kbps. Synchronization of the interferometer is performed by OTES using the spacecraft clock. The OTES power interface uses redundant unregulated power lines to the input of the

OTES DC/DC converter. The unregulated input power can be from +24 to +34 VDC, with a nominal value of 28 VDC.

*2.3.6 OTES Radiation and Contamination Mitigation*

The OTES is based on the MGS TES and Mini-TES instruments that were designed, tested, and operated for >5 years in a radiation environment very similar to that of OSIRIS-REx. All previous instruments were designed to survive 20 krad environments. It is estimated that the MGS TES exceeded this radiation dose in its 10-year life, with no detectable radiation damage effects. The OTES was built following the Goddard Space Flight Center and OSIRIS-REx Project assembly and test procedures to minimize any organic contamination carried on the instrument. Contamination on the instrument would have to be extreme (>10s of μm) to degrade the performance based on experience with Mini-TES in the dusty surface environment of Mars (Ruff et al. 2011). Analysis and monitoring of witness samples through the OTES development at ASU demonstrate the OTES was delivered with a cleanliness level D per NASA CR 4740.

*2.3.7 OTES Operational Modes*

OTES is a point spectrometer with a single detector that begins collecting telemetry data when power is applied. Upon receipt of the data collection command, OTES collects and transmits interferograms to the spacecraft continuously when operational. Global maps of Bennu based on OTES data will be constructed through one-dimensional slewing of the spacecraft to point the OTES across the asteroid surface from north to south and back continuously over one asteroid rotation period. This slew pattern, together with Bennu's rotation, results in near-complete coverage of the asteroid surface. Maps of potential sample sites will be compiled from data obtained during low-altitude reconnaissance passes over the asteroid surface, combined with cross-track slewing of the spacecraft. OTES has two operational modes—data collection and standby. In data collection mode OTES collects one interferogram every two seconds; in standby mode no interferogram data are collected, but telemetry data are collected once every two seconds. Instrument commands are used to insert the internal cal mirror into the optical path and to switch between the redundant fringe laser diodes and start-of-scan optical switches.

### 2.3.8 Commanding

The OTES is commanded from the spacecraft using discrete, time-tagged commands. The OTES data collection command syntax is: OTES_ACQ *sclk osclk acq_id slews dwell1 gain1 dwell2 gain2 dwell3 gain3*

where:

| | |
|---|---|
| *sclk* | Spacecraft clock, S/C command transmit time |
| *osclk* | Spacecraft clock, OTES command execution time |
| *acq_id* | Acquisition identifier |
| *slews* | Number of slews in acquisition (repeat loop) |
| *dwell1* | Number of interferograms of scene zone 1 (per acquisition) |
| *gain1* | Scene zone 1 gain (low/high) |
| *dwell2* | Number of interferograms of scene zone 2 (per acquisition) |
| *gain2* | Scene zone 2 gain (low/high) |
| *dwell3* | Number of interferograms of scene zone 3 (per acquisition) |
| *gain3* | Scene zone 3 gain (low/high) |

The OTES calibration data collection command syntax is: OTES_CAL *sclk osclk cal_id cal_dwell cal_gain*

where:

| | |
|---|---|
| *sclk* | Spacecraft clock, S/C command transmit time |
| *osclk* | Spacecraft clock, OTES command execution time |
| *cal_id* | Calibration identifier |
| *cal_dwell* | Number of interferograms of calibration flag |
| *cal_gain* | Cal gain (low/high) |

## 2.4 Interferometer

OTES Michelson interferometer uses a linear moving mirror to collect double-sided interferograms in both the IR and a 0.855-µm laser metrology interferometer that share a common beamsplitter. Each time the moving mirror physically moves one half of the laser wavelength, the metrology laser signal goes to zero and the IR signal is sampled. The spectral sampling is a function of the total mechanical path length on one side of the interferogram, which is in turn determined by the laser wavelength and the number of interferogram samples collected. Initially, OTES was configured to collect 1186 ±3 double-sided interferogram samples. After delivery the servo velocity and sampling frequency were increased to mitigate a sample jitter noise induced by the spacecraft inertial measure unit (IMU) [see Section 5]. Following this adjustment, OTES collects 1350 ±3 samples. In order to have a consistent spectral sampling accounting for any sampling change over the life of the mission, the interferogram is filled with zeros to a uniform length of 1360 samples prior to performing the discrete Fourier transform. This number of samples, together with the measured laser wavelength [see Section 6.4] results in a spectral sampling of 8.66 cm$^{-1}$.

# 3 Calibration Methodology

## 3.1 Calibrated Radiance

In an ideal Michelson interferometer a beamsplitter sends half of the energy from the external scene down each of the two arms of the interferometer, where it reflects off of either a fixed or a moving mirror. Phase shifts of 180° occur at each mirror reflection, and 90° at each beamsplitter reflection. The energy from the two arms recombines at the beamsplitter, where half of the energy is reflected toward the detector and half is transmitted through the beamsplitter and exits out the telescope aperture. At the zero path difference location, and each time the moving mirror moves one half a wavelength λ (d = n × ½λ; n ≥ 0), the waves reaching the detector are in phase and combine in constructive interference, whereas the waves exiting the interferometer are 180° out of phase and combine destructively (Griffiths and deHaseth 2007). As a result, at path differences of ½nλ, all of the external energy reaches the detector and none of the energy exits the interferometer. The energy emitted from the detector follows the opposite path, and at path differences of ½nλ, all of the energy from the detector exits out through the instrument aperture and none returns to the detector. At each ¼λ path difference (d = (¼ + ½n)λ), the situation is reversed, and the waves from the source combine at the detector in destructive interference and the energy emitted from the detector returns to the detector in constructive interference. As a result, an interferometer measures the difference in the spectral radiance (in W cm$^{-2}$ sr$^{-1}$ /cm$^{-1}$) coming alternatively from the external scene and from the detector itself.

The external radiance is the sum of the radiance from the scene through the aperture as limited by the field stop ($I_{scene}$), the radiance from the optics themselves ($I_{optics}$), and any radiance from the field stop and the interior of the instrument that is modulated through the interferometer ($I_{instrument}$). The difference between this modulated external radiance and that from the detector ($I_{detector}$), each weighted by their viewing solid angle, is the signal measured by the detector. This formulation can be generalized to include non-isothermal conditions, the emission and transmission of each of the optical elements, and the specifics of the optical design and calibration methodology.

Using these definitions, the electrical signal ($V$) produced by the interferometer is given by:

$$V = \left(I_{scene} + I_{optics} + I_{instrument} - I_{detector}\right) IRF \quad (1)$$

where:

$IRF$ = instrument response function in converting the optical input to electrical output

Observations of calibration targets allow the $I$ and $IRF$ terms to be determined and used to determine the scene radiance.

The radiance from the scene is given by:

$$I_{scene} = \varepsilon_{scene} B_{scene} \tau_{optics} \quad (2)$$

where:

$\varepsilon_{scene}$ = emissivity of the scene

$B_{scene}$ = Planck emission from the scene

$\tau_{optics}$ = throughput of the full optics

The radiance from the full optics ($I_{optics}$) is the sum of the emitted radiance from each element, attenuated by the reflectivity of the remaining optical elements along the optical path to the detector. This radiance is given by:

$$I_{optics} = \sum_{i=1}^{nelements} \varepsilon_i B_i \tau_{i+1} \quad (3)$$

where:

$n_{elements}$ = total number of optical elements

$\varepsilon_i$ = emissivity of optical element i

$B_i$ = Planck radiance from optical element i

$\tau_{i+1}$ = throughput of the optical elements downstream from element i given by:

$$\tau_{i+1} = \rho_{i+1} * \rho_{i+2} * ... * \rho_{nmirrors}$$

$\rho_i$ = reflectivity (or transmissivity) of optical element i

If all the elements have the same reflectivity, $\rho$, then:

$$I_{optics} = \left(\varepsilon_1 B_1 \rho^{n-1}\right) + \left(\varepsilon_2 B_2 \rho^{n-2}\right) + \left(\varepsilon_3 B_3 \rho^{n-3}\right) ... + \left(\varepsilon_n B_n \rho^{n-n}\right) \quad (4)$$

Replacing $\varepsilon$ with its equivalent $(1-\rho)$ gives:

$$I_{optics} = \left(B_1(1-\rho_1)\rho^{n-1}\right) + \left(B_2(1-\rho_2)\rho^{n-2}\right) + \left(B_3(1-\rho_3)\rho^{n-3}\right) ... + \left(B_n(1-\rho_n)\rho^{n-n}\right) \quad (5)$$

The optics radiance ($I_{optics}$) can be divided into the emitted radiance from the fore and aft optics. In an interferometer the aft optics radiance does not pass through the interferometer and is not modulated, so only the fore optics radiance becomes a component of the interferogram signal. The radiance from the instrument ($I_{instrument}$) includes the radiance emitted from the field stop, which is modulated through the interferometer. In practice, however, this modulated radiance will be the same in all observations taken within a short period of time and is indistinguishable from the detector radiance; this radiance will be included in the detector radiance in the subsequent discussion. With these refinements Eq. 1 becomes:

$$V = \left(I_{scene} + I_{fore} - I_{detector}\right) IRF \quad (6)$$

Expanding these terms gives:

$$I_{scene} = \varepsilon_{scene} B_{scene} \tau_{optics} \quad (7a)$$

$$I_{fore} = \varepsilon_{primary} B_{primary} \rho_{secondary} + \varepsilon_{secondary} B_{secondary} \quad (7b)$$

$$I_{detector} = \varepsilon_D B_D \quad (7c)$$

The throughput of the optics ($\tau_{optics}$) in Eq. 7a is given by:

$$\tau_{optics} = \tau_{fore} \tau_{aft} \quad (8a)$$

where for OTES:

$$\tau_{fore} = \rho_{primary} \rho_{secondary} \quad (8b)$$

$$\tau_{aft} = \rho_1 \rho_2 \cdots \rho_n = \rho^n \quad (8c)$$

where $\rho$ is the reflectivity of each element, and $n$ is the number of elements in the aft optics. The determination of the desired scene radiance ($I_{scene}$) from Eq. 6 requires knowledge of the two unknown terms ($I_{fore} - I_{detector}$) and the *IRF*, which in turn requires the observation of two calibration targets of known radiance, both, one, or neither of which is viewed through the identical optical path as the scene. In the following discussion it is assumed that at least one calibration target is observed through the same optical path as the target (full optics calibration); in many cases this target is space, but it can be any target whose temperature and emissivity are known. For OTES the internal calibration blackbody is located behind the fore optics, and the *IRF* cannot be determined simply from observations of the two calibration targets because the same optical path is not viewed in all three observations (scene, space, and cal blackbody), and the $I_{fore}$ term does not cancel as it does in the two-target, full-optics method. In this case the fore optics throughput and emission must be estimated. The uncertainties in these estimates result in errors in the scene radiance, the magnitude of which can be investigated by modeling the scene radiance as the optical throughput and emission are varied over their uncertainties.

The measured spectra, *V*, from each observation are given by Eq. 6:

$$V_{scene} = (I_{scene} + I_{fore} - I_{detector}) IRF \quad (9a)$$

$$V_{space} = (I_{space} + I_{fore} - I_{detector}) IRF \quad (9b)$$

$$V_{cal} = (I_{cal} - I_{detector}) IRF \quad (9c)$$

The instrument response function, *IRF*, can be computed by differencing the cal and space spectra:

$$V_{cal} - V_{space} = ((I_{cal} - I_{detector}) - (I_{space} + I_{fore} - I_{detector})) IRF \quad (10)$$

$$V_{cal} - V_{space} = (I_{cal} - I_{fore} - I_{space}) IRF$$

giving:

$$IRF = \frac{V_{cal} - V_{space}}{I_{cal} - I_{fore} - I_{space}} \quad (11)$$

The terms in this equation can be expanded to give:

$$I_{cal} = \varepsilon_{cal} B_{cal} \rho_{flag} \tau_{aft} + \varepsilon_{flag} B_{flag} \tau_{aft} \quad (12a)$$

$$I_{space} = \varepsilon_{space} B_{space} \tau_{fore} \tau_{aft} \quad (12b)$$

$$I_{fore} = \left(\left(\varepsilon_{primary} B_{primary}\right)\rho_{secondary} + \varepsilon_{secondary} B_{secondary}\right)\tau_{aft} \quad (12c)$$

Substituting into Eq. 11 gives:

$$IRF = \frac{V_{cal} - V_{space}}{\tau_{aft}\left(\varepsilon_{cal} B_{cal} \rho_{flag} + \varepsilon_{flag} B_{flag}\right) - \left(\varepsilon_{primary} B_{primary} \rho_{secondary} + \varepsilon_{secondary} B_{secondary}\right)\tau_{aft} - \varepsilon_{space} B_{space}\left(\tau_{fore} \tau_{aft}\right)} \quad (13)$$

or

$$\tau_{aft} IRF = \frac{V_{cal} - V_{space}}{\varepsilon_{cal} B_{cal} \rho_{flag} + \varepsilon_{flag} B_{flag} - \left(\varepsilon_{primary} B_{primary} \rho_{secondary} + \varepsilon_{secondary} B_{secondary}\right) - \varepsilon_{space} B_{space} \tau_{fore}} \quad (14)$$

Because both the space and scene observations are acquired through the full optics, Eq. 9a and 9b can be differenced to give:

$$V_{scene} - V_{space} = \left(I_{scene} - I_{space}\right) IRF \quad (15)$$

The calibrated scene radiance is then given by:

$$I_{scene} = \frac{V_{scene} - V_{space}}{IRF} + I_{space} \quad (16)$$

Expanding gives:

$$I_{scene} = \varepsilon_{scene} B_{scene} \tau_{fore} \tau_{aft} = \frac{V_{scene} - V_{space}}{IRF} + \varepsilon_{space} B_{space} \tau_{fore} \tau_{aft}$$

$$\varepsilon_{scene}B_{scene} = \frac{V_{scene} - V_{space}}{\tau_{fore}\left(\tau_{aft}IRF\right)} + \varepsilon_{space}B_{space} \quad (17)$$

Substituting for ($\tau_{aft}IRF$) gives:

$$\varepsilon_{scene}B_{scene} = \frac{V_{scene} - V_{space}}{\tau_{fore}\left(\frac{V_{cal} - V_{space}}{\left[\varepsilon_{cal}B_{cal}\rho_{flag} + \varepsilon_{flag}B_{flag} - \left(\varepsilon_{primary}B_{primary}\rho_{seconday} + \varepsilon_{primary}B_{secondary}\right) - \varepsilon_{space}B_{space}\tau_{fore}\right]}\right)} + \varepsilon_{space}B_{space}$$

or:

$$\varepsilon_{scene}B_{scene} = \left(\frac{V_{scene} - V_{space}}{V_{cal} - V_{space}}\right)\left(\frac{\varepsilon_{cal}B_{cal}\rho_{flag} + \varepsilon_{flag}B_{flag} - \left(\varepsilon_{primary}B_{primary}\rho_{secondary} + \varepsilon_{secondary}B_{secondary}\right)}{\tau_{fore}} - \varepsilon_{space}B_{space}\right) + \varepsilon_{space}B_{space} \quad (18)$$

## 3.2 Absolute Calibration

The absolute accuracy of the scene radiance was modeled by assessing the effects of the uncertainties in each of the key variables in Eq. 18 that are used to compute the calibrated radiance. This equation has been applied in a Monte Carlo model to assess the absolute error in the scene radiance using: (1) the random noise errors in the measured signals ($V_{scene}$, $V_{space}$, and $V_{cal}$); (2) the measurement errors in the radiance from the cal and space targets ($B_{cal}$, and $B_{space}$) and the primary and secondary mirrors ($B_{primary}$, $B_{secondary}$) that are due to temperature errors in the measurement of each of these elements; and (3) the errors in the emissivity of the calibration blackbody and the reflectivity of the primary and secondary mirrors and the calibration flag. The measurement errors in the measured spectra were modeled using the instrument response function and the NESR of the Flight OTES instrument [see Section 6].

The OTES simulated spectra were converted into calibrated scene radiance that was then converted to brightness temperature at each wavenumber, assuming an emissivity of unity. The surface kinetic temperature, which is the desired parameter for temperature mapping, was determined by averaging the brightness temperature over a specified wavenumber range. This process is roughly equivalent to fitting a blackbody radiance curve to the measured radiance to provide a best-fit determination of the surface temperature. The use of multiple spectral samples to estimate the surface temperature reduces the uncertainty in the temperature measurement from the instrument noise (NESR).

Figure 5 shows the radiance data for 40 representative spectra for a scene temperature of 325 K converted to brightness temperature as a function of wavenumber, illustrating the nature of the random and systematic errors. Figure 6 gives the derived surface temperature determined by averaging the brightness temperatures from 300 to 1200 cm$^{-1}$ for 500 cases. The 1-sigma error in an individual OTES temperature measurement can be determined by taking the standard deviation of the derived surface temperatures in Figure 6.

[Insert Figure 5]

[Insert Figure 6]

Table 4 gives the modeled absolute errors viewing a 300 K blackbody using the nominal values for each variable, along with the expected worst-case uncertainties, which were then varied both individually and collectively using a Monte Carlo approach to determine the integrated radiance. For each individual case the number of random trials computed was 10,000; for the full case of all parameters varying, the number of trials was 100,000.

[Insert Table 4]

The temperature and emissivity of the cal target are the largest contributors to error. The total RMS radiance error from this analysis, for realistic values of the OTES temperature, emissivity, and reflectivity knowledge, is 0.91%, which meets the OTES requirement.

The uncertainty in the derived surface temperature as a function of surface temperature was determined using a similar analysis. In this modeling the measured OTES Flight instrument NESR [see Fig. 20a] was combined with the worst-case expected instrument parameter uncertainties given in Table 4 in a Monte Carlo model that created simulated calibrated radiance spectra using the method outlined above. In this model the brightness temperature was again computed at each wavenumber, but because the surface temperature was varied from 70 K to 400 K, the wavenumber range over which the surface temperature was determined was varied with temperature. The full spectral range was used for warm targets, whereas narrower low-wavenumber ranges were used at lower surface temperatures. As a result, the NESR begins to affect the accuracy of the derived surface kinetic temperature at temperatures below ~125 K because fewer spectral channels are available to average the brightness temperature. The model was run for 10,000 random cases for surface temperatures from 70 to 400 K at 10 K intervals, and the standard deviation of the derived surface kinetic temperature was computed for each surface temperature (Fig. 7). These values represent the best estimate of the OTES absolute temperature accuracy errors. As seen in Figure 7 OTES readily meets the absolute temperature requirement of ±5 K at 70 K and ±2 K for 220 to 350 K surfaces.

[Insert Figure 7]

# 4    Experimental Setup

## 4.1    Test Equipment and Facilities

The OTES instrument was assembled, tested, and calibrated on the Arizona State University Tempe campus in an ISO Class 7 (class 10,000) cleanroom in building ISTB4, with Class 6 (1,000) flow benches for component assembly. The OTES calibration was done in both ambient and thermal vacuum conditions. All test equipment was calibrated to NIST standards on a routine basis. Bench-level testing of the OTES instrument consisted of piece-part and system-level testing of each subsection under ambient conditions, followed by system performance testing that was conducted before, during, and after each environmental and thermal-vacuum test. Ambient tests determined the field-of-view definition and alignment, the out-of-field response (encircled energy), the spectrometer spectral sample position and spectral range, the ambient instrument functional performance, and verified the command and data links. The thermal vacuum calibration testing determined the signal gain values, the emissivity and temperature of the internal calibration blackbody, the instrument response function and its variation with instrument temperature, and the radiometric precision and accuracy. Throughout the integration and test phase an Instrument Functional Test (IFT) was performed using the OTES ambient target and two calibration reference blackbody standards (Bench Checkout Units; BCU) (Fig. 8). This test provided the baseline functional and limited performance assessment for the instrument and was performed before, during, and after environmental tests in order to demonstrate that the instrument functionality was not degraded by the tests. A Comprehensive Performance Test (CPT) was performed using the ambient and BCU targets and provided the baseline radiometric performance assessment for the instrument. This test was also performed before, during, and after environmental tests. The OTES optical field of view was characterized using an 8-inch-diameter f/7.5 off-axis collimator that was scanned in elevation and azimuth using computer-controlled actuators.

[Insert Figure 8]

The OTES thermal vacuum radiometric calibration was performed using two external calibration targets whose absolute temperature was determined using precision thermistors that were NIST-calibrated to ±0.1 °C. The emissivity of these targets was determined to be 0.99897 ±0.0002 based on the spectral properties of the PT-401 paint and the analysis of a cone blackbody (Bedford 1988; Prokhorov et al. 2009). The temperature uncertainly of the optics was determined using the NIST calibration of the flight thermistors mounted to these elements, and the reflectivity was determined using laboratory measurements of the gold coatings that were applied to the optical surfaces.

OTES was radiometrically calibrated in vacuum at instrument temperatures of –10, 0, 10, 25, and 40 °C. Calibration tests were performed by viewing the two precision BCUs, one set at a temperature of 85 K to simulate space observations and a second set to temperatures of 150, 190, 220, 260, 300, 340, and 380 K to span the expected Bennu surface temperatures. OTES was installed in a LACO thermal vacuum chamber developed for the OTES project in the Class 8 (100,000) cleanroom in ISTB4 on the ASU campus (Fig. 9). OTES was mounted on a metallic L-plate with thermostatically controlled strip heaters that simulated the spacecraft nadir deck for thermal balance and was used to drive the instrument temperature during thermal vacuum testing and thermal cycling. An LN2 cold plate was attached to the back of the mounting plate to provide a cooling source for the bracket/instrument. OTES was configured in the chamber to allow its aperture to point between the two NIST-calibrated external blackbody calibration targets using a vacuum-compatible rotary stage attached to the underside of the L-plate (Fig. 9). The instrument and mounting bracket were fully blanketed with flight-like Multi-Layer Insulation (MLI) thermal blankets to simulate the spacecraft thermal configuration. Ground Support Equipment (GSE) controlled the OTES, L-plate, targets, and rotary stage via cabling through the chamber feed-throughs.

[Insert Figure 9]

### 4.2 Software/Scripts

The OTES data were processed using the open source vector math software package called *davinci* that was developed at ASU for instrument data processing and has been used for in-flight calibration and processing of TES, Mini-TES, and THEMIS data. The OTES performance varies slightly between the forward and reverse moving mirror scan directions, so all processing is done separately for the two scan directions. The discrete Fourier transform (DFT) is performed on the interferograms, bad spectra are identified and removed, and the internal and external calibration targets and scene target are identified using either telemetry, the rotary table viewing angle, or user-supplied input files. Groups of calibration spectra are averaged and interpolated, the radiometric calibration is performed using the methods described in Section 3.1, and the analysis of the data is performed.

## 5 OTES Development

### 5.1 Flight Instrument Development

The OTES instrument began development in May 2011 with the selection of the OSIRIS-REx mission for Phase B. The System Requirement Review (SRR) was held in December 2011, and the System Definition Review in March 2012. The Preliminary Design Review (PDR) was held on Nov. 27–29, 2012, with the Critical Design Review (CDR) 11 months later on Dec. 4–6, 2013. The OTES was delivered to the spacecraft 18 months after CDR on June 25, 2015, with a total development time from SRR to delivery of 42 months. Few design changes occurred between PDR and CDR, and no significant changes were made following CDR. The most challenging development in the OTES program was the design, fabrication, and coating of the diamond beamsplitter. This element was the largest precision optical-quality diamond yet produced by Diamond Materials, and the antireflection microstructure (ARM) etching on the beamsplitter was the largest and most uniform etch yet produced by TelAztec. The ARM process necessary to meet the size and uniformity requirements took several iterations to perfect, but did not affect either the overall OTES cost or development schedule.

OTES underwent thermal vacuum radiometric testing from May 21 through June 15, 2015. These tests included radiometric calibration at multiple instrument temperatures, as well as six thermal cycles from −15 to +45 °C (Fig. 10). During the radiometric testing, it was determined that the absolute calibration was not meeting the expected performance. Extensive analysis suggested that the data at all instrument and target temperatures could be fit if it was assumed that the liquid-nitrogen-temperature BCU-1 unit was 9–11 °C warmer than the thermistors indicated, and if the warm target (BCU-2) temperature was 0 to +2 °C different than the thermistors indicated and varied with instrument temperature. These differences could be best explained by a delamination of the paint on the interior blackbody cone surface, in which the radiating surface of the paint was at a different temperature than the aluminum body of the target in which the thermistors are embedded. The modeled temperature and emissivity differences are consistent with a first-order modeling of the temperature gradients between the radiating paint surface and the underlying aluminum in a vacuum, and paint delamination explains why the best-fit effective temperature of the paint varies systematically with the instrument temperature.

[Insert Figure 10]

The Project schedule requirements did not permit the repair of the BCUs and a repeat of the Flight instrument radiometric testing. Therefore, the effects of the paint delamination will be investigated using the OTES Engineering Model (EM). The EM will be fitted with the flight spare electronics, detector, and telescope, and a flight-like interferometer and aft optics, and the thermal vacuum radiometric testing will be repeated. The first tests will use the BCU blackbodies with the delaminated paint and confirm the problems observed with the flight instrument and provide a baseline. The BCUs will then be repainted and a subset of the tests will be repeated. The intent of this testing is to verify that the delamination of the paint was the cause of the radiometric errors that were observed with the Flight instrument. The primary ramification of the BCU issue is that the transfer of the NIST calibration from the BCUs to the internal calibration target was not as accurate as desired. The recovery from this shortcoming is discussed in Section 6.5.

## 5.2    Post-Delivery Testing

Following the delivery of the Flight instrument to the spacecraft developer and flight system integrator, Lockheed Martin Space Systems (LMSS; Bierhaus et al. 2017), three noise sources were observed during spacecraft testing. The first consisted of extremely high noise on the IR signal that occurred without any degradation in the servo performance. This noise source was traced to audio noise that was produced by a sonic orifice that was used to regulate the N2 gaseous purge rate to OTES. Pyroelectric detectors are sensitive to audio microphonics, and the cause of the noise was confirmed by reducing or eliminating the N2 purge rate, at which time the OTES noise disappeared. A final confirmation was accomplished by successfully reproducing the magnitude and frequency characteristics of the noise using the EM.

A second noise source is related to vibrations induced by the spacecraft reaction wheels when they were operated at >4,000 rpm. Wheel speeds of this magnitude are not expected in flight, and if they were to occur their only effect would be to produce a single, narrow noise spike in the OTES spectrum.

The third noise source is associated with the spacecraft inertial measurement units (IMU). These units produce discrete vibrations in the 500–700 Hz range that have a magnitude of ~5–10 milli-G at the OTES spacecraft mounting surface. In its original configuration, the OTES IR interferometer collected samples at 656 Hz. The servo control is designed to fit a sine wave to the metrology laser signal and predict when the laser signal will go to zero and IR signal should be sampled (Section 2.3). Unfortunately the IMU-induced vibrations result in small, periodic jitter in the moving mirror position, and the zero-position of the mirror can have slight (~1%) offsets. The magnitude of these offsets are small (0.01 × ½ × 0.849 μm or 0.004 μm), but these offsets are periodic in nature and result in artificial sinusoids at specific frequencies in the OTES IR signal. These frequencies correspond to the difference in frequency between the OTES 656 Hz sampling frequency and the IMU frequencies, and occur at ~150, 100, and 40 Hz. The OTES information band is from 10 to 120 Hz, so only the latter two frequencies produced noise in the OTES IR spectrum. This problem was resolved by increasing the servo velocity to 0.321 mm/s, giving a sampling frequency of 772 Hz. The resulting frequency differences are ~300, 200, and 150, all of which are outside the OTES information band. The increase in servo velocity resulted in more IR spectral bands and a smaller spectral sampling interval, but does result in a higher noise. This effect is discussed in Section 6.7; the values given in Table 3 correspond to the higher servo velocity that will be used in flight.

# 6    Pre-Launch Calibration Results

## 6.1    Gain

The OTES gains were set to nominal values of 1×, 2×, and 4× using individual precision resistors. Gain 2× was set for the nominal instrument performance and expected Bennu temperatures. Gains of 1× and 4× are designed for unexpected behaviors or asteroid characteristic and for possible degradations. The actual gain values were determined using data acquired during the thermal vacuum testing at five instrument (detector) temperatures (Table 5). The method used the interferogram peak-to-peak (P-P) value, averaged for 100 spectra in both the forward and reverse scan directions at each of the three gain states for each instrument temperature. The peak-to-peak values correspond to the ZPD position in the interferogram and have the highest signal-to-noise ratio, providing the best indicator of the true gain values. For each test the target temperature was ~27 °C. Table 5 gives the average and standard deviation of the P-P for each test condition, along with the ratio of the P-P values for Gain 2×/1× and Gain 4×/1×. As seen in this table there is no systematic temperature variation in the gain values. As a result the gain value is assumed to be the average of all measurements at all temperatures. The values in Table 5 are within 1-σ of precisely 2 and 4, respectively, and the gains will be assumed to be precisely 2 and 4.

[Insert Table 5]

## 6.2    Field of View

The OTES field of view was determined in azimuth and elevation using the collimator viewing a glowbar target through 1-mrad-wide vertical or horizontal slits placed at the collimator focus. The collimator was oriented parallel to the OTES optical axis (0 azimuth, 0 elevation). A thermally stable shutter was used that was alternatively opened to view the glowbar/slit, or closed to view the shutter in order to remove any thermal drift in the system. Twenty spectra were collected with the shutter open and 20 with the shutter closed at each azimuth or elevation step. The slit was moved at 0.5-mrad steps, starting and ending well outside of the nominal field of view. The analysis averaged the value of the peak-to-peak of the ZPD position of the interferogram in order to maximize the signal-to-noise ratio. The thermal drift was removed, and data were normalized to the minimum and maximum P-P values. The field of view results following the environmental testing and final pre-ship alignments are given in Figure 11. Small negative values occur near the edge of field of view that are due to small thermal gradients at the edges of the slit. These small artifacts do not have a significant effect on the derived field of view response, and the zero response level in the signal normalization was set to be at locations where the slit edge effects were not present. The field of view, defined at the full-width, half-maximum (FWHM) point in the normalized data, is 6.47 mrad in azimuth and 6.22 mrad in elevation. The OTES optical axis was measured relative to the OTES alignment cube, and a transformation matrix for the optical axis relative to the spacecraft-mounting interface was provided to LMSS for the OTES spacecraft alignment and pointing determination.

[Insert Figure 11]

### 6.3 Encircled Energy

The OTES encircled energy performance was determined using the collimator with a set of fixed apertures with diameters of 2, 4, 6, 8, 12, 16, 20, 28, and 36 mrad that were placed at the collimator focus. The encircled energy test collected 30 spectra with the shutter open and 30 with the shutter closed at each azimuth or elevation step. The analysis used the averaged values of peak-to-peak of the interferogram to maximize the signal-to-noise ratio. The thermal drift was removed, and data were normalized to the minimum and maximum P-P values. The results are shown in Figure 12, and demonstrate that OTES easily exceeds its requirement for 85% of the measured energy coming from one geometric scene footprint of 6.47 mrad.

[Insert Figure 12]

### 6.4 Spectral Sample Position and Spectral Range

The OTES spectral sample position is a function of the total Michelson mirror displacement, which is determined by the metrology laser wavelength and the number of samples that were collected in each interferogram. The nominal OTES laser wavelengths were 0.8561 μm and 0.8553 μm for laser 1 and laser 2, respectively, at 25 °C, and 0.8528 and 0.8526 μm for laser 1 and laser 2, respectively, at –20 °C. However, the two lasers were slightly off the optical axis, and because the divergence of the IR beam is greater than the laser beam there is a small wavelength shift from the nominal laser wavelength values (Griffiths and deHaseth 2007). This shift is linear with wavenumber and can be determined using measurements of a known spectral target (Griffiths and deHaseth 2007). During testing a polyethylene target was observed using the primary laser (laser 1) under ambient conditions. The results from this test found the best-fit value for the laser wavelength to be 0.849 µm. Spectra of the lab atmosphere containing $CO_2$ and water vapor were collected using both laser 1 and 2 and were found to be identical. A value of 0.849 will be used for both the primary and redundant lasers. Over the full expected OTES operational temperature, the laser wavelength only varies by $<\pm0.17\%$, and the resulting OTES IR sample spacing by $<\pm0.015$ cm$^{-1}$, and the laser wavelength will be assumed to be constant over temperature.

## 6.5    Internal Calibration Blackbody Properties

The OTES internal calibration blackbody is used for absolute calibration, and the knowledge of its temperature and emissivity are key to the absolute accuracy. This blackbody target is a 1-cm-diameter, 15° half-angle cone that has two precision thermistors installed in its walls (Fig. 13). These thermistors were calibrated to an absolute accuracy of 0.1 K, and the interior surface was painted with PT-401 paint that has an emissivity of >0.92. The energy from this target reflects off of a gold-plated mirror (Fig. 13) toward the interferometer. This cal mirror rotates 60° from its stowed position to reflect the radiance from the blackbody to the interferometer. The cal mirror is not instrumented directly, but it is stowed in the cal mirror cavity, whose temperature is determined from redundant, precision thermistors (Fig. 13). The cal mirror also provides Sun protection during launch and spacecraft safing events.

[Insert Figure 13]

The actual temperature and emissivity of the calibration blackbody were determined in testing by transferring the NIST calibration accuracy of the BCU external blackbodies to the internal blackbody. The temperature of the internal blackbody was determined in thermal vacuum testing using the two external BCU blackbodies in Eq. 18 and treating the internal target as an unknown. The internal blackbody emissivity requirement is an emissivity of >0.98 across the performance wavenumber range of 300 to 1350, with a 1-sigma uncertainty of <0.005. Figure 14(a) gives the emissivity of the internal calibration blackbody derived from the best available measurements. Figure 14(b) gives the 1-sigma variation in the measurements used to derive the emissivity and indicates that OTES meets the requirements, except for a small discrepancy at wavenumbers >1300 cm$^{-1}$. However, as discussed in Section 5.1, these data are uncertain due to the issues associated with the paint on the BCUs. Data will be collected in 2017 using the OTES EM under vacuum conditions that will be used to provide a better estimate of the true emissivity and the uncertainty in the internal calibration blackbody. This test will use the flight spare electronics, telescope, and an interferometer. Most importantly, the internal calibration blackbody used in this test was fabricated, painted, instrumented, and assembled at the same time as the flight unit using flight spare components, and its emissivity and temperature precision and accuracy should be nearly identical to the flight OTES. The emissivity of this calibration blackbody assembly will be determined by transferring the NIST calibration from the BCU targets. If the data from this assembly do not have the issues seen when calibrating the flight unit, then they will provide a more accurate estimate of the properties of the flight unit and will be used for the in-flight calibration.

[Insert Figure 14]

The internal blackbody temperature uncertainty requirement is 1 °C with a precision of ±0.2 °C. Figure 15 gives the measured and derived values of the internal calibration blackbody. Because the blackbody temperature was determined by fitting the Planck function to the calibrated radiance, the temperatures are not sensitive to the small errors in emissivity, and OTES meets its calibration target temperature requirement.

[Insert Figure 15]

## 6.6 Instrument Response Function

The OTES Instrument Response Function (IRF) provides the transfer from photons into the instrument to the signal (voltage) output from the instrument as an interferogram. The IRF varies with wavenumber due to the wavelength variations of the detector response, the diamond beamsplitter and lenses, and all of the beamsplitter and mirror finishes and coatings. The IRF was determined as a function of instrument temperature using one of the BCU calibration blackbody standards and the internal calibration blackbody during the thermal vacuum radiometric calibration. Figure 16(a) gives an example of the IRF for an instrument temperature of 10 °C, at the original servo velocity and IR sampling frequency of 656 Hz. The variation of IRF with instrument temperature for a representative set of wavenumbers is given in Figure 16(b). The primary cause of variation in response function is the variation in detector performance with temperature, and to a lesser extent, changes in the interferometer alignment with temperature. Over the OTES in-specification operational temperature range of 10 to 40 °C and the performance spectral range of 300 to 1350 cm$^{-1}$, the OTES IRF only varies by <11%. As a result, the OTES radiometric noise performance will vary by 11% over the full, expected operational environment.

[Insert Figure 16]

The data in Figure 16 were collected during pre-delivery thermal vacuum testing at ASU with the original IR sampling frequency of 656 Hz. Unfortunately, no precision radiometric data could be collected with the Flight unit after the IR sampling frequency was increased to 772 Hz [see Section 5.2]. The only expected change to the IRF would be potential minor changes due to the pyroelectric detector frequency response. These potential changes were investigated using the EM, which has a detector from the same lot as the Flight unit. Figure 17 shows the comparison of the EM unit at the 656- and 772-Hz sampling frequencies. The signal values are different from the Flight unit because of different gain settings, and the spectral signatures of $CO_2$ and water vapor are present, so these data cannot be directly compared to the Flight unit data in Figure 16(a). However, the key result in Figure 17 is that the IRF does not change significantly with an increase in servo velocity. Because of the uncertainties in the testing done with the EM at ambient conditions—the target temperature was less stable than during the thermal vacuum radiometric testing, and both $CO_2$ and water vapor are present and potentially varying in the path—the small differences seen in the IRFs in Figure 17 are likely due to the test conditions rather than to real differences in the IRF with servo velocity. Therefore, we conclude that the IRF data presented in Figure 15 are the same for the Flight unit at the 772-Hz sampling frequency. This conclusion has been verified in flight during cruise using observations of space and the internal blackbody.

[Insert Figure 17]

## 6.7 Precision: Noise Equivalent Spectral Radiance

The OTES precision, or noise equivalent spectral radiance (NESR), was determined by calculating a scene radiance using the two precision BCU calibration targets, together with Eq. 18. The NESR is defined as the standard deviation (1-sigma) of the scene radiance and was determined during thermal vacuum testing operational instrument temperatures of 10, 20, and 40 °C. In addition to the intrinsic noise in the system, any changes in the temperature of the target or the instrument during the radiometric test will result in changes in the scene radiance derived from Eq. 18, which then results in an NESR that is larger than the true system noise. In order to minimize these additional noise sources, the OTES instrument and the BCU targets were maintained at as stable a temperature as possible. Temperature changes in the instrument and the targets are inherent when alternatively viewing calibration targets at different temperatures. This occurs because, despite active thermal control, the radiative coupling between the instrument and the targets acts to drive their temperatures toward a common value, resulting in thermal drifts that, while small (<<0.1 °C/minute), are still large compared to the NESR. In order to minimize these drifts, the duration of the data collection was limited to 200 s at each target.

The NESR results are given in Figure 18. These data were collected during pre-delivery thermal vacuum testing at ASU with the original moving mirror velocity and an IR sampling frequency of 656 Hz. The data presented are the average of the NESR for instrument temperatures of −10, 0, 10, 20, and 40 °C, gains of 2× and 4×, and viewing both the external and internal calibration targets. The normal operating gain of OTES is 2×; a gain of 1× is included as a safety factor for any unexpected changes in the instrument or in the scene temperature, but it does not fully digitize the OTES noise. There are very minor differences in the measured spectra collected when the Michelson moving mirror is moving in the forward direction when compared to spectra collected in the reverse direction. Therefore, the most accurate measurement of the NESR is computed processing the forward and reverse data separately. For data collected in flight the forward and reverse direction data are processed independently to remove any artifacts that might be introduced.

[Insert Figure 18]

At the original sampling frequency OTES meets the performance requirement over the full range from 300 to 1350 cm$^{-1}$ (Fig. 18). However, as discussed in Section 5.2, the moving mirror velocity and IR sampling frequency were increased to move the observed microphonic interference from the IMU out of the OTES information band. A complete and systematic set of data could not be collected to determine the OTES NESR after the velocity increase. However, data collected with the Flight unit during thermal vacuum testing at LMSS were used to investigate the magnitude of the increase in the NESR at the higher servo velocity. These data results are shown in Figure 19 and are for data collected using the internal blackbody, whose temperature was very stable, that were converted to NESR using the instrument response function. The OTES instrument performance model predicts a 20% increase in the NESR at the higher servo velocity. Figure 19 gives the comparison of the pre-delivery NESR at 656 Hz with the measured NESR at 772 Hz, along with the 656-Hz NESR scaled by 20%, and the OTES requirement. As seen in this figure, the 772-Hz NESR is very nearly 20% higher at the higher velocity over most of the spectral range as expected, with a few minor deviations that are associated with vacuum chamber and instrumentation noise during the LMSS tests. Based on these results we conclude that the pre-delivery NESR, increased by 20%, provides the best estimate of the NESR at a sampling frequency of 772 Hz. These data are given in Figure 20(a). Even at the 772-Hz IR sampling frequency, OTES meets the NESR requirement, except for a single spectral channel at the extreme high-wavenumber end. The corresponding signal-to-noise ratio is given in Figure 20(b) at the reference temperature of 325 K.

[Insert Figure 19]
[Insert Figure 20]

## 6.8  Absolute Accuracy

The OTES absolute calibration accuracy determination was complicated by the paint delamination issues discussed in Section 5.2. However, analysis of the spectral shape of the derived calibrated radiance indicated that there were systematic offsets in the effective radiative temperature of the warm (BCU2) inner cone surface relative to the temperatures indicated by the embedded precision thermistors. For an instrument temperature of 10 °C, these temperature offsets are 0.5, 1.25, 1.25, 1.25, 1.5, and 2 K for target temperatures of 148.94, 188.46, 218.75, 259.44, 299.83, and 380.53 K, respectively. Figure 21(a) shows the calculated OTES calibrated radiance for the BCU, and the blackbody radiance using the measured BCU2 temperature with the estimated temperature offsets for an instrument temperature of 10 °C. Figure 21(b) shows the absolute value of the difference between these radiances; comparison with Figure 18 shows that the absolute accuracy is at most only approximately four times larger than the NESR. The integrated radiance error from 6 to 50 µm is given in Table 6, and the results meet the OTES Yarkovsky requirement.

[Insert Figure 21]
[Insert Table 6]

## 6.9  Linearity

The OTES linearity requirement is 10%. The OTES performance was determined using the thermal vacuum radiometric data collected at instrument temperatures of 10, 20, and 40 °C, which cover the OTES performance in specification requirement range. The measured voltage spectra viewing the internal calibration blackbody and the external BCU were integrated from 100 to 1350 cm$^{-1}$ to produce a delta signal. The radiance from the internal blackbody was calculated using its measured temperature, and the calibrated radiance from the external BCU was calculated using Eq. 18. These radiances were differenced to produce a delta radiance. These data are shown in Figure 22 for all of the calibration tests conducted at 10, 20, and 40 °C instrument temperatures. A linear function was fit to the data (Fig. 22). This function goes through zero, as expected for a linear, well-calibrated instrument, and the maximum deviation of the data from this linear function is <3.7%, and is well below this value for all but the lowest delta radiances, and readily meets the OTES requirement.

[Insert Figure 22]

# 7  The OTES Mission Operations

## 7.1  In-Flight Operation

OTES will be operated using spacecraft-commanded sequences. An OTES sequence consists of power on, a warm-up and thermal stabilization period, a series of alternating observations of the external scene and the internal calibration blackbody, and power off. At power on, the OTES moving mirror begins operating, and OTES begins collecting and outputting telemetry information. In this "stand-by" mode, no interferogram data are output. An OTES command from the spacecraft puts the instrument into its interferogram output "science" mode. In this science mode OTES collects groups of interferograms viewing the external scene with the calibration flag open, and groups of interferograms viewing the internal calibration blackbody with the calibration flag closed. During the Detailed Survey period, the spacecraft will be slewed back and forth across Bennu, with the slews designed to take the OTES field of view well beyond Bennu to acquire observations of space. In this manner, space, cal, and scene interferograms are collected and used to compute the Bennu calibrated radiance. For other periods in which OTES is not slewed periodically to observe space, the space observations at the beginning and end of the sequence will be used if available. If no space views are available, then a previously determined Instrument Response Function will be used.

## 7.2  In-Flight Calibration

The OTES instrument data will be calibrated in flight using Eq. 18, together with periodic views of space and the internal calibration blackbody. Typical operation will collect 10 interferograms of the internal calibration blackbody and ~10 interferograms of space during each spacecraft slew across the asteroid. All interferograms are converted to spectra through a discrete Fourier transform. Following this transform, the spectra in each group of space and cal observations will be averaged and then interpolated between successive space or calibration observations. If only a single group of space or cal observations are collected, these will be used for the entire sequence. If science spectra are collected before or after the first or last space or cal observation, these bounding space or cal observations will be extended forward or backward in time as appropriate. If only calibration blackbody or space observations are acquired in a sequence, then the pre-launch IRF, at the appropriate instrument temperature (Fig. 15(b)), will be used with the available calibration observation in Eq. 16 to determine the scene radiance. If no calibration observations are available, then the detector thermistor temperature will be used to estimate $I_{detector}$, and will be used with the IRF to estimate the scene radiance using Eq. 9a. These one-point and "zero-point" calibration results would only be used if an anomaly or failure occurred, but would be less accurate than the normal two-point calibration using both space and the internal blackbody. The magnitude of this uncertainty will be determined in flight if necessary. In-flight experience with the TES and Mini-TES instruments demonstrated that the instrument response function varied by less than 5% over the life of these missions (Christensen et al. 2001, 2004b), and a similar stability is expected for the OTES instrument. Changes in the instrument temperature during a sequence require periodic observations of the internal calibration blackbody. In-flight experience with the TES instrument (Christensen et al. 2001), together with the OTES thermal modeling, indicate that the instrument temperature will vary by <0.1 °C/minute, and that these temperature variations are nearly linear. As a result, the internal blackbody will typically be observed every 5–10 minutes, and these observations will be linearly interpolated. With experience at Bennu if a more complex function is needed to account for the variations in instrument temperature, it will be implemented.

# 8     Data Processing and Archiving

The OTES data processing pipeline (Fig. 23) is straightforward based on the flight-proven TES and Mini-TES methodology and software. Each OTES sequence, from power on to power off, will be processed independently, and the interferograms in the forward and reverse scan directions will be processed separately. Observations of space will be identified using the spacecraft pointing and viewing data and the Science Processing and Operations Center (SPOC) software (Lauretta et al. 2017). Each interferogram will be transformed to a spectrum. The periodic observations of space and the internal blackbody will be used to produce a calibrated radiance spectrum for each observation. Bad or noisy spectra will be identified and removed prior to averaging groups of space or calibration data. These averaged space and cal spectra will then be interpolated or extrapolated to provide a space and cal spectrum ($V_{space}$ and $V_{cal}$) for each scene spectrum. The temperatures of the internal blackbody and optical elements will be determined using the redundant thermistor data, with testing done to eliminate erroneous data. These data provide the inputs to Eq. 18 that are used to compute the calibrated radiance for each scene spectrum.

[Insert Figure 23]

The OTES standard data product is calibrated radiance. These data, along with all of data used in the calibration, will be delivered to the Planetary Data System on the schedule agreed to by the OSIRIS-REx Project. For each spectrum the calibrated radiance is converted to brightness temperature at each wavenumber, assuming unity emissivity. The processing continues to compute surface temperature in a full emissivity-temperature separation process and mineral abundance using the emissivity with a spectral library. Temperature uncertainties are computed for use in thermal inertia error analyses and landing site selection. These uncertainties are computed using an initial estimate of the surface temperature from the brightness temperature to constrain the temperature-dependent wavenumber range over which to average the brightness temperature, computing a new surface temperature, and repeating until the results converge. The resulting temperature is termed the average brightness temperature, which is then used with Figure 7 to compute the brightness temperature uncertainty.

Additional data products that will utilize OTES data will include the separation of the calibrated radiance into surface temperature and emissivity, mineral abundance determinations, dust cover index, and thermal inertia.

## Acknowledgments

We would like to thank Mary Walker, Jim Hendershot, and their team for their excellent support in the OSIRIS-REx Payload Office, Sam Pellicori for his engineering support, Tara Fisher and Ashley Toland for priceless administrative support, and Selex-Galileo, Diamond Materials, General Dynamics, Avior, BEI, TelAztec, and L&M Machining. Ed Cloutis provided a very helpful review. This material is based upon work supported by the National Aeronautics and Space Administration under contract NNM10AA11C issued through the New Frontiers Program.

# Figure Captions

**Fig. 1** Representative meteorite and mineral spectra acquired at the OTES spectral resolution

**Fig. 2** The OTES CAD model, in an expanded view, showing the major elements of the OTES modular design

**Fig. 3** The OTES block diagram, showing the key functional elements and the instrument-to-spacecraft interfaces

**Fig. 4 (a)** The OTES CAD model showing the interferometer layout on the aft optics plate. **(b)** The OTES optical ray trace, showing the telescope, aft optics, interferometer fixed and moving mirrors, IR detector, and internal calibration blackbody

**Fig. 5** Brightness temperature versus wavenumber for a representative suite of random instrument temperature and emissivity errors

**Fig. 6** Random error in derived surface temperature. Each value is the average of the brightness temperature over a specified wavenumber (Fig. 5) for 500 representative random instrument errors (see text for details)

**Fig. 7** OTES absolute temperature uncertainty

**Fig. 8 (a)** The CAD model of the OTES external calibration targets (BCU1 and BCU2) and the OTES instrument in their configuration for thermal vacuum radiometric testing. **(b)** The BCU targets and the Flight OTES installed on the thermal control plate and rotary table in the ASU thermal vacuum chamber

**Fig. 9** OTES Flight instrument and BCU calibration targets in the thermal vacuum chamber on the ASU campus immediately prior to the start of thermal vacuum testing

**Fig. 10** The OTES thermal balance and thermal vacuum test timeline

**Fig. 11** The OTES field of view measured following environmental and thermal vacuum testing, and immediately prior to delivery to the spacecraft. The small negative values at the edge of the measured field of view are due to small temperature gradients in the 1-mrad slit used to map the field of view and do not significantly affect the results. **(a)** Azimuth. **(b)** Elevation

**Fig. 12** The OTES encircled energy measured following environmental and thermal vacuum testing, and immediately prior to delivery to the spacecraft. The purple data point shows the requirement of ≥85% of the energy in an 8-mrad field, which OTES easily achieves

**Fig. 13** The OTES CAD model showing the design and layout of the internal calibration blackbody, with the cal flag assembly, which consists of an Avior bi-stable rotary actuator that moves a small flag with a gold-coated mirror, the calibration blackbody with redundant thermistors, and the cal flag cavity with redundant thermistors that stores the cal flag mirror when not in use. The cal flag is closed when OTES is powered off for Sun protection

**Fig. 14** The OTES internal calibration blackbody properties. These data were collected during the Flight OTES thermal vacuum radiometric calibration. **(a)** Internal calibration blackbody emissivity. The paint on the external BCU blackbodies delaminated slighting during the testing, so these data are likely the lower limit of the Flight

OTES blackbody emissivity (see text). Data collected with the OTES EM under vacuum conditions will be used to better determine the true emissivity of this target. **(b)** The 1-sigma uncertainty in the OTES internal calibration blackbody emissivity. These data are not affected by the BCU target issues

**Fig. 15** The measured and derived temperature of the internal calibration blackbody. The measured temperature is the average of the two precision thermistors mounted on the exterior of the internal blackbody [see Fig. 13]

**Fig. 16 (a)** The OTES Instrument Response Function (IRF) for an instrument temperature of 10 °C. **(b)** The variation in the OTES Instrument Response Function (IRF) with temperature at five representative wavenumbers. These data are all of the data collected at the different instrument temperatures and target temperatures during thermal vacuum radiometric testing

**Fig. 17** Comparison of the Instrument Response Function (IRF) at IR sampling frequencies of 656 Hz and 772 Hz. These data were collected using the EM unit and are representative of the expected performance of the Flight unit. The small differences are within the uncertainties in these ambient measurements, and we conclude that the IRF data for the Flight unit in Fig. 16 do represent the IRF at the 772-Hz IR sampling frequency

**Fig. 18** The OTES noise equivalent delta radiance (NESR), which represents the 1-sigma variation in calibrated radiance. These data were collected in thermal vacuum radiometric testing at the original moving mirror sampling frequency of 656 Hz

**Fig. 19** Comparison of the noise equivalent delta radiance (NESR) at 656 Hz and 772 Hz. The 772-Hz sampling frequency data were collected using the Flight model during thermal vacuum testing at LMSS. Also shown are the pre-delivery, 656-Hz sampling frequency data, scaled by 20%, and the OTES requirement

**Fig. 20 (a)** The OTES noise equivalent delta radiance (NESR), scaled by the modeled and measured increase in NESR of 20% at a sampling frequency of 772 Hz. These data provide the best estimate of the NESR for the Flight unit at the 772 Hz IR sampling frequency. **(b)** The OTES signal-to-noise ratio (SNR), based on the data from Fig. 20a, for a scene temperature of 325 K

**Fig. 21 (a)** The OTES calibrated radiance for the BCU, along with the blackbody radiance computed using the BCU thermistor temperatures with the delta temperatures discussed in the text applied. These data were acquired at an instrument temperature of 10 °C. **(b)** The absolute value of the difference between the calibrated radiances given in (a) and those of an ideal blackbody at the measured BCU thermistor temperature

**Fig. 22** The OTES linearity. Delta radiance is the difference between the calibrated radiance from the external BCU, calculated using Eq. 18, and the radiance from the internal blackbody, calculated using its measured temperature, integrated from 100 to 1350 $cm^{-1}$, in units of W $cm^{-2}$ $sr^{-1}$ /$cm^{-1}$. The delta signal is the difference between the measured spectrum ($V_{scene}$) at $V_{cal}$, again integrated from 100 to 1350 $cm^{-1}$ in units of transformed volts

**Fig. 23** The OTES data processing pipeline

# Tables

**Table 1** Comparison of ASU infrared instruments

| Instrument | Spectral Range | Spectral Resolution | # Bands | IFOV (mrad) | Mass (kg) | Power (W) | Size (cm) |
|---|---|---|---|---|---|---|---|
| MO/MGS TES | 148–1651 cm$^{-1}$ | 9.80; 4.90 cm$^{-1}$ | 143 | 8.3 | 14.4 | 14.5 | 23.6 × 35.5 × 40.0 |
| MER Mini-TES | 340–1998 cm$^{-1}$ | 9.99 cm$^{-1}$ | 167 | 20 | 2.4 | 5.6 | 23.5 × 16.3 × 15.5 |
| Odyssey THEMIS | 6.3–14.5 μm | ~1 μm | 10 | 0.25 | 11.2 | 14.0 | 29 × 38 × 55 |
| OTES | 100–1750 cm$^{-1}$ | 8.66 cm$^{-1}$ | 193 | 6.5 | 6.27 | 10.8 | 37.5 × 28.9 × 52.2 |

**Table 2** Yarkovsky absolute accuracy requirement

| Scene Temperature (K) | Integrated Radiance Requirement (1.5%) |
|---|---|
| 220 | $6.819 \times 10^{-6}$ W cm$^{-2}$ sr$^{-1}$ |
| 300 | $2.359 \times 10^{-5}$ W cm$^{-2}$ sr$^{-1}$ |
| 350 | $4.211 \times 10^{-5}$ W cm$^{-2}$ sr$^{-1}$ |

**Table 3** OTES as-built instrument parameters

| Parameter | Value |
|---|---|
| Spectral range | 1750 to 100 cm$^{-1}$ (5.71–100 μm) |
| Spectral resolution | 8.66 cm$^{-1}$ |
| Telescope aperture | 15.2 cm |
| f# | 3.91 |
| Field of view (FWHM) | 6.47 mrad in azimuth and 6.22 mrad in elevation |
| Detector | uncooled deuterated L-alanine doped triglycine sulfate (DLATGS) pyroelectric |
| Detector D* | 1.2 10$^9$ cm Hz$^{1/2}$ watt$^{-1}$ at 10 Hz, 22 °C |
| NESR | 1.73 × 10$^{-8}$ W cm$^{-2}$ str$^{-1}$/cm$^{-1}$ at 300 cm$^{-1}$<br>1.72 × 10$^{-8}$ W cm$^{-2}$ str$^{-1}$/cm$^{-1}$ at 1000 cm$^{-1}$<br>2.17 × 10$^{-8}$ W cm$^{-2}$ str$^{-1}$/cm$^{-1}$ at 1350 cm$^{-1}$ |
| Cycle time per measurement | 1.8 s plus 0.2 s scan reversal time |
| Metrology laser nominal wavelength | 25 °C: laser 1: 0.8561 μm; laser 2: 0.8553 μm<br>–20 °C: laser 1: 0.8528 μm; laser 2: 0.8526 μm |
| Metrology laser self-apodized wavelength (25 °C) | laser 1: 0.849 μm; laser 2: 0.849 μm |
| Michelson mirror travel | ±0.289 mm |
| Michelson mirror velocity | 0.321 mm/s |
| Sampling frequency | 772 Hz |
| Number of bits per sample | 16 |
| Number of samples per interferogram—Nominal | 1350 ±3 |
| Number of samples per interferogram—Filled | 1360 |
| Nominal data volume per 2 s interval | data: 11,312 bits; telemetry: 200 bits |
| In-flight calibration | two-point; internal calibration blackbody and space |
| Cal blackbody emissivity | 0.98 ±0.005 |
| Thermal requirements | performance in specification: 10 °C to 40 C<br>flight allowable operational range: –15 °C to +45 °C<br>nonoperational proto-flight survival range: –25 °C to +55 °C |
| Solar protection | cal mirror in stowed position |
| Mass | 6.27 kg |
| Power | 10.8 W average; 15.99 W peak |
| Dimensions | 37.49 × 28.91 × 52.19 cm |

**Table 4** Model absolute radiance percentage error

| Variable | Nominal Value | Uncertainty | % Integrated Radiance Error (6–50 μm) |
|---|---|---|---|
| T cal target | 10 °C | 0.5 °C | 0.73% |
| ε cal target | 0.99 | 0.005 | 0.54% |
| T flag | 10 °C | 1 °C | 0.034% |
| T primary/secondary | 10 °C | 0.75 °C | 0.035% |
| R primary/secondary | 0.985 | 0.005 | 0.027% |
| Monte Carlo of All (Requirement = 1.5%) | | | 0.91% |

**Table 5** Average and standard deviation of the P-P at five instrument temperatures

| Detector Temperature (°C) | –4.9 | 3.6 | 13.4 | 29.8 | 44.3 | Average |
|---|---|---|---|---|---|---|
| Ave: P-P: 1× | 0.06949 | 0.07882 | 0.08789 | 0.09260 | 0.08929 | — |
| σ: P-P:1× | 0.0005258 | 0.0005188 | 0.0006877 | 0.0005793 | 0.0004977 | — |
| Ave: P-P: 2× | 0.1392 | 0.1580 | 0.1757 | 0.1857 | 0.1785 | — |
| σ: P-P:2× | 0.0006185 | 0.0006174 | 0.0008908 | 0.0006334 | 0.0005446 | — |
| Ave: P-P: 4× | 0.2787 | 0.3168 | 0.3521 | 0.3732 | 0.3574 | — |
| σ: P-P:4× | 0.001462 | 0.001301 | 0.001780 | 0.0009511 | 0.0009298 | — |
| Ave: Gain 2×/1× | 2.0026 | 2.0046 | 1.9989 | 2.0057 | 1.9989 | 2.0021 |
| σ: Gain 2×/1× | 0.00962 | 0.0107 | 0.0100 | 0.00950 | 0.00855 | 0.010028 |
| Ave: Gain 4×/1× | 4.0125 | 4.0179 | 4.0076 | 4.0254 | 4.0028 | 4.0076 |
| σ: Gain 4×/1× | 0.0214 | 0.0167 | 0.0204 | 0.0173 | 0.0128 | 0.010891 |

Table 6. OTES estimated accuracy error

| Target Temperature | Integrated radiance (6–50 μm) error (%) |
|---|---|
| 148.94 | 0.278 |
| 188.45 | 1.05 |
| 218.75 | −0.053 |
| 259.44 | −0.728 |
| 299.83 | 0.213 |
| 380.53 | 0.108 |

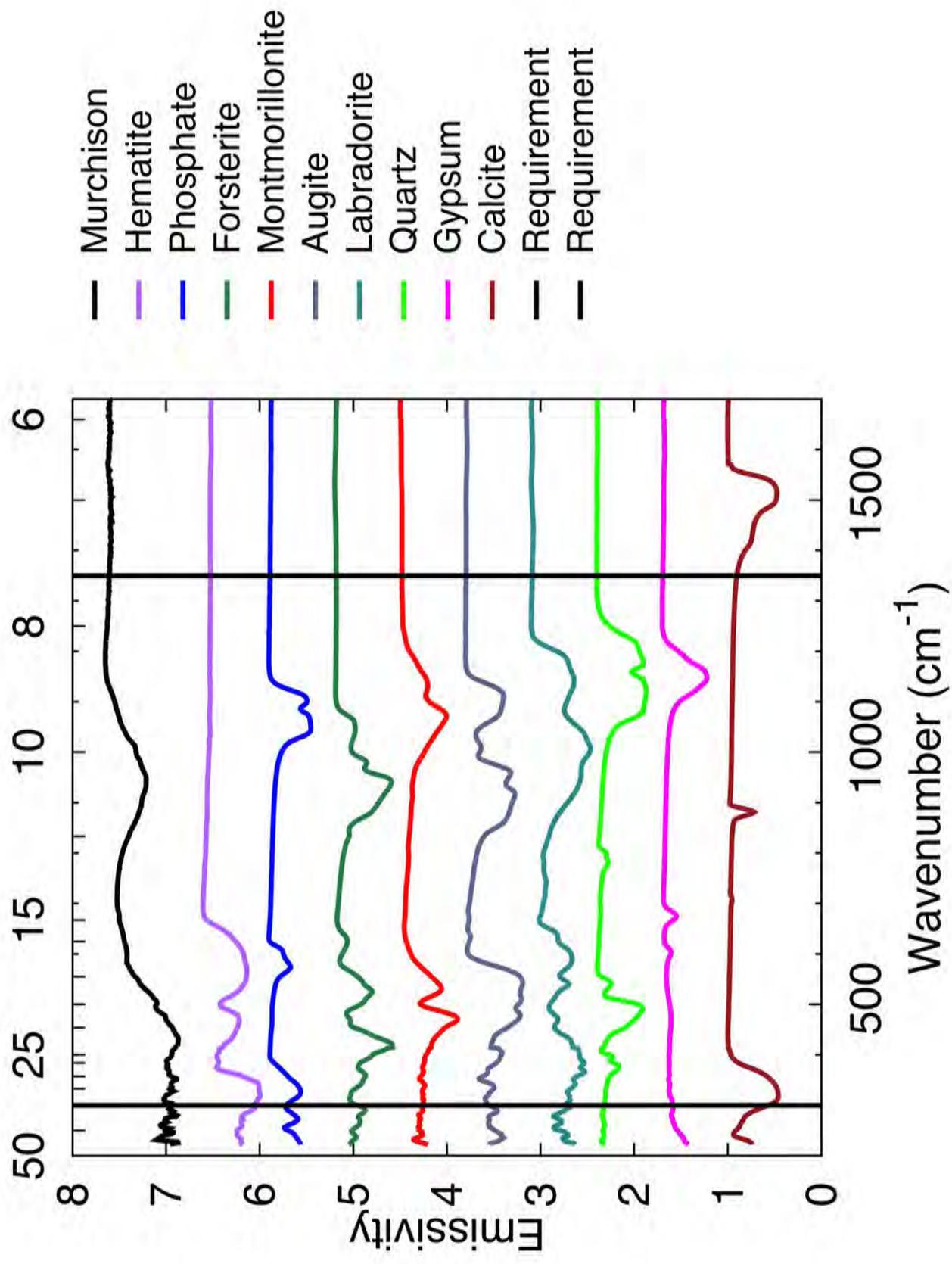

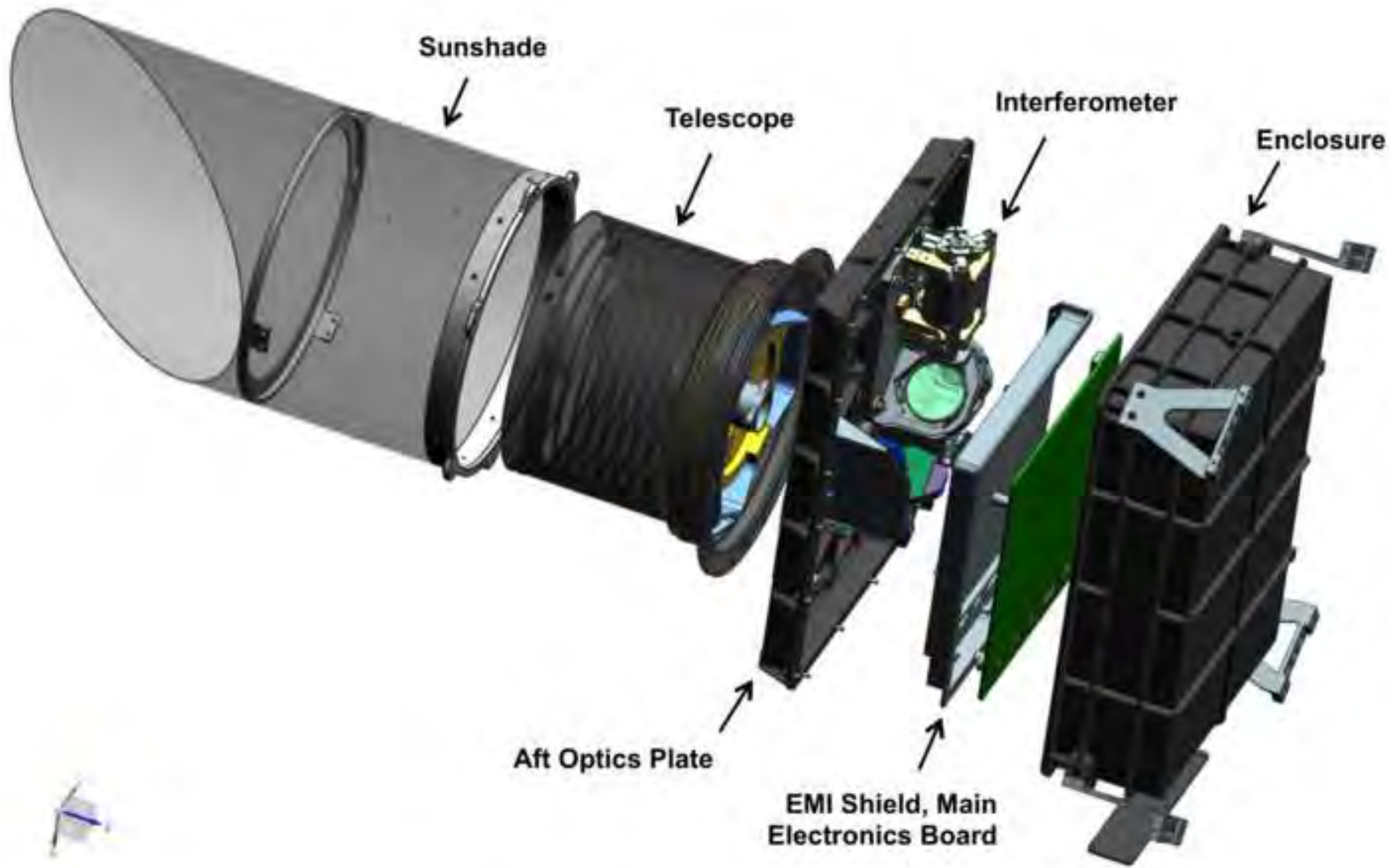

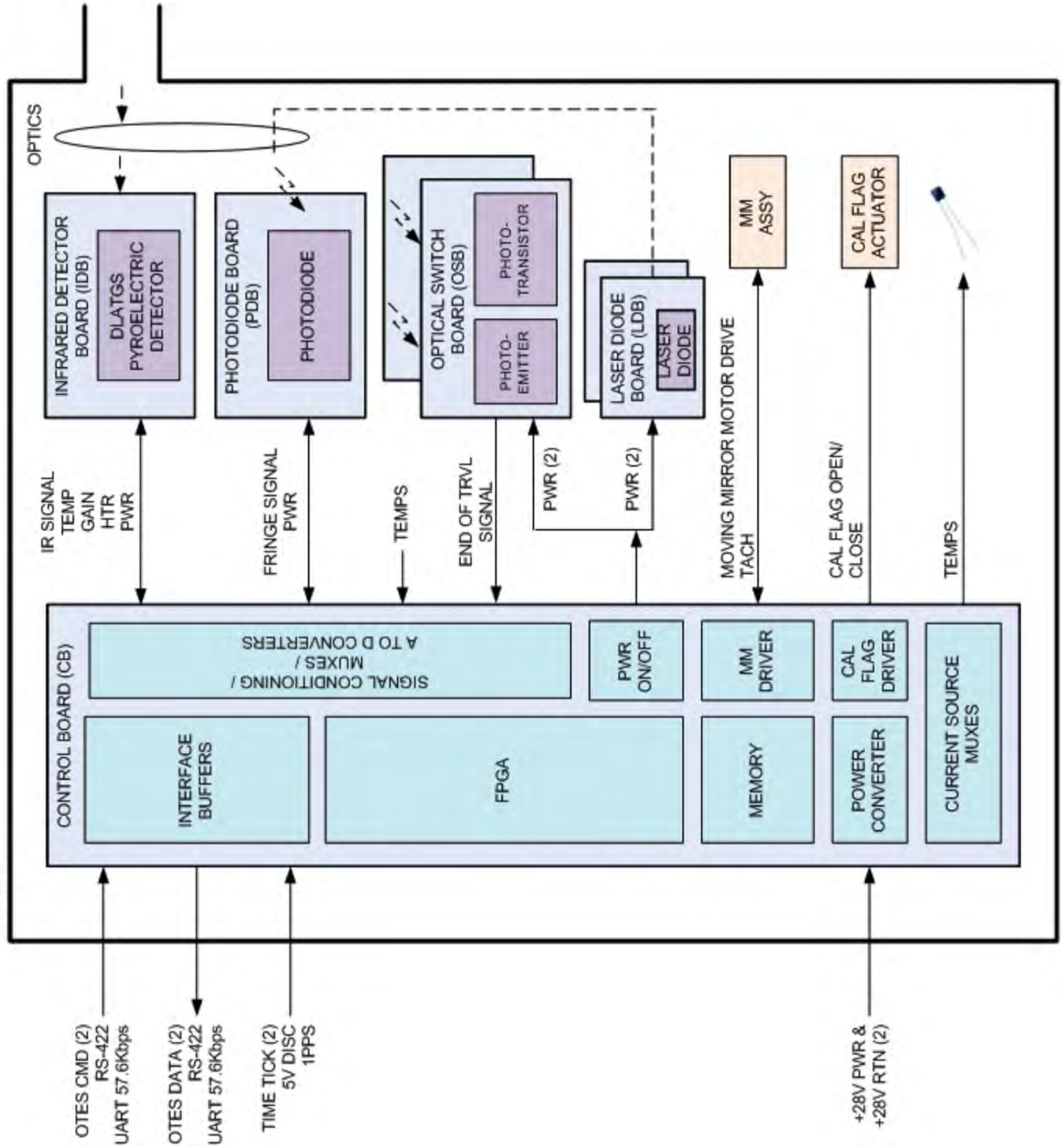

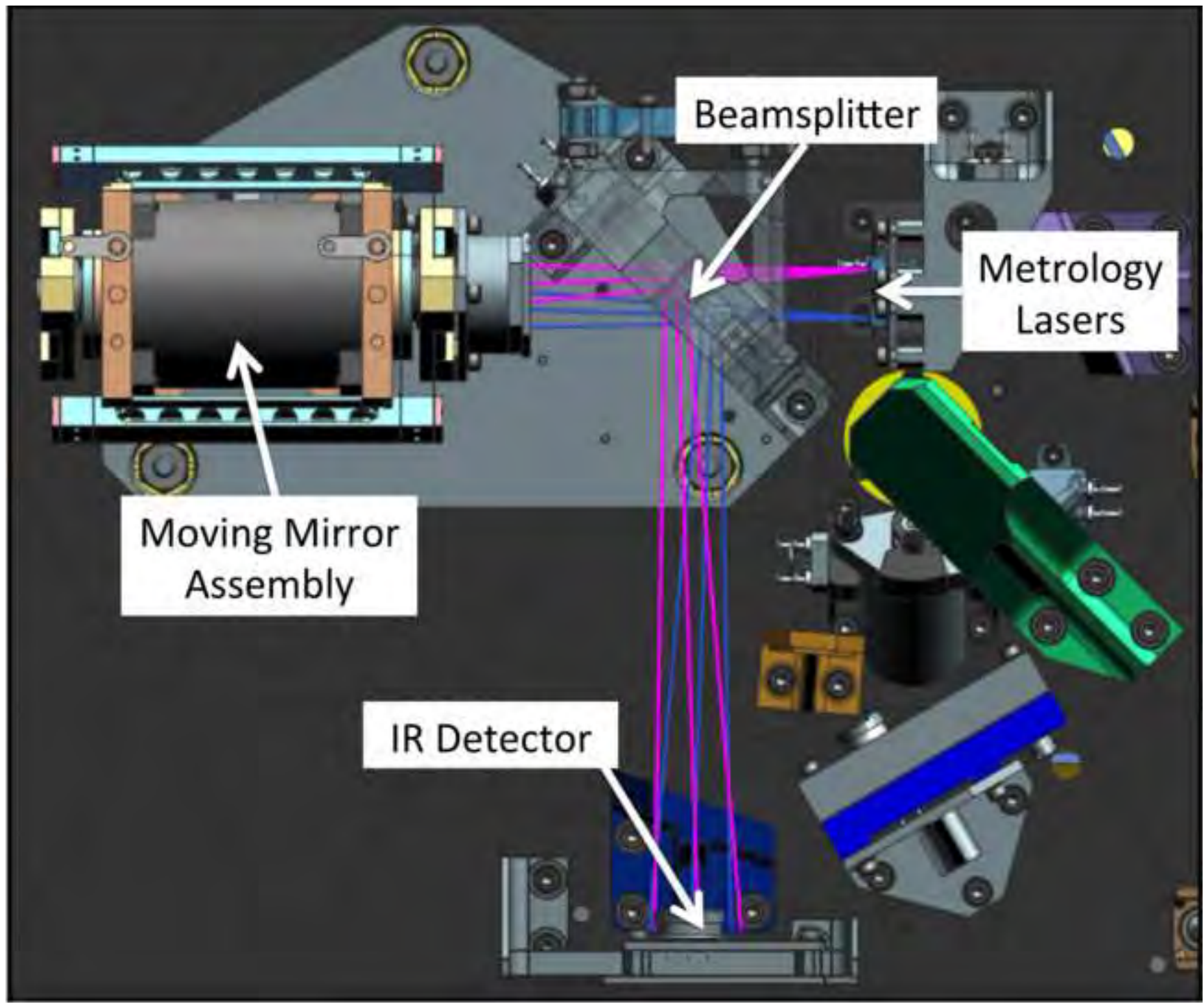

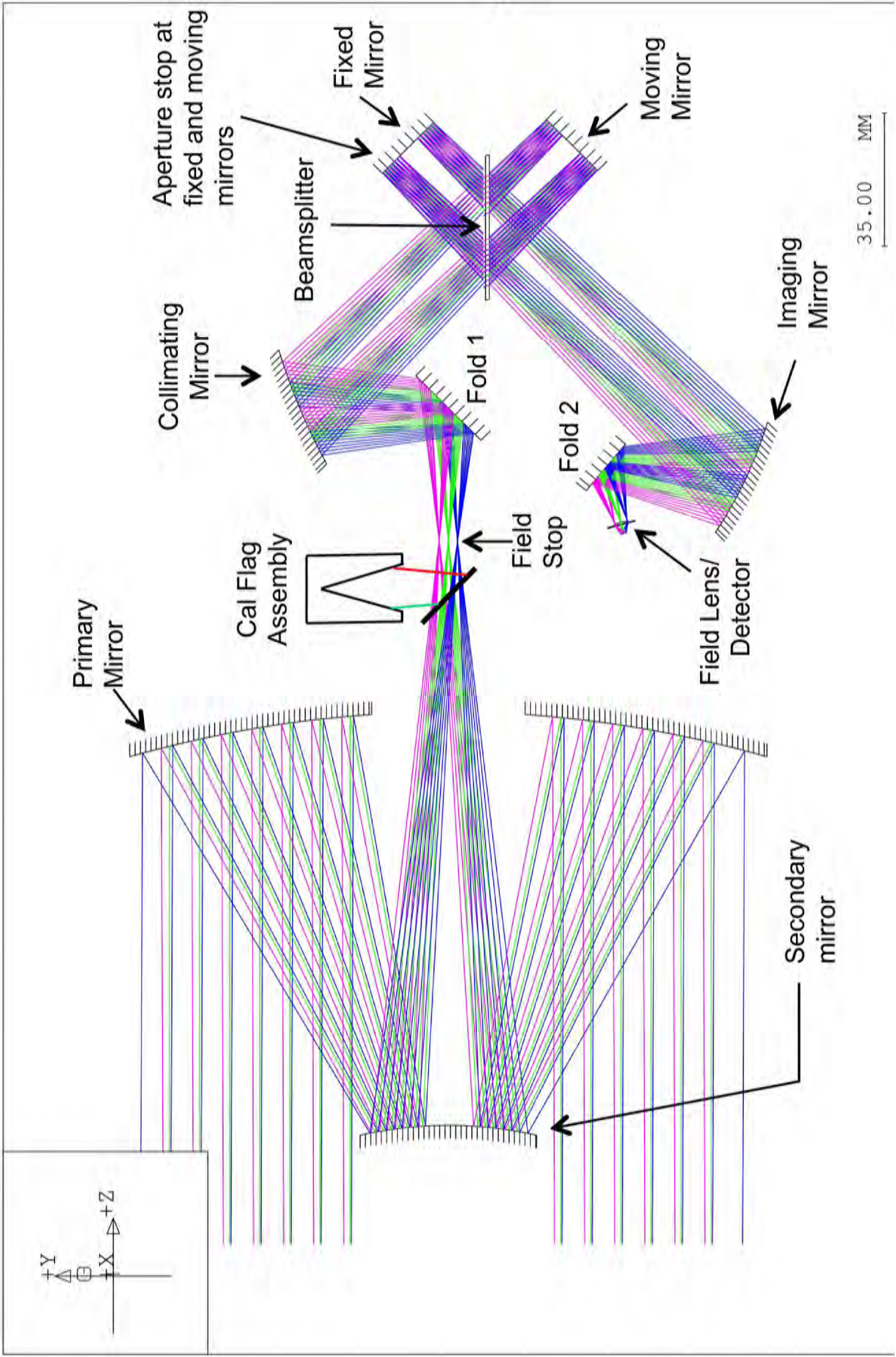

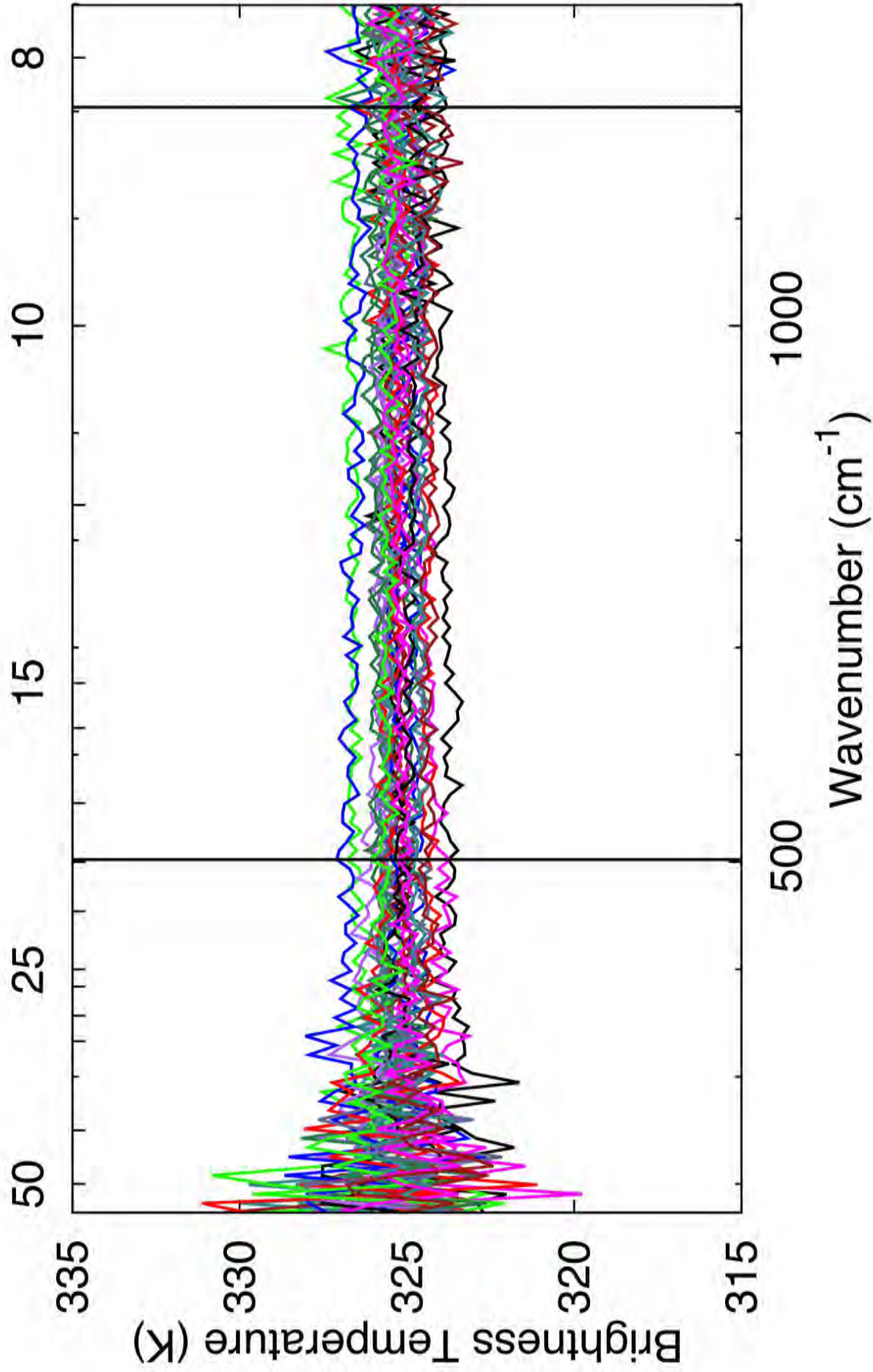

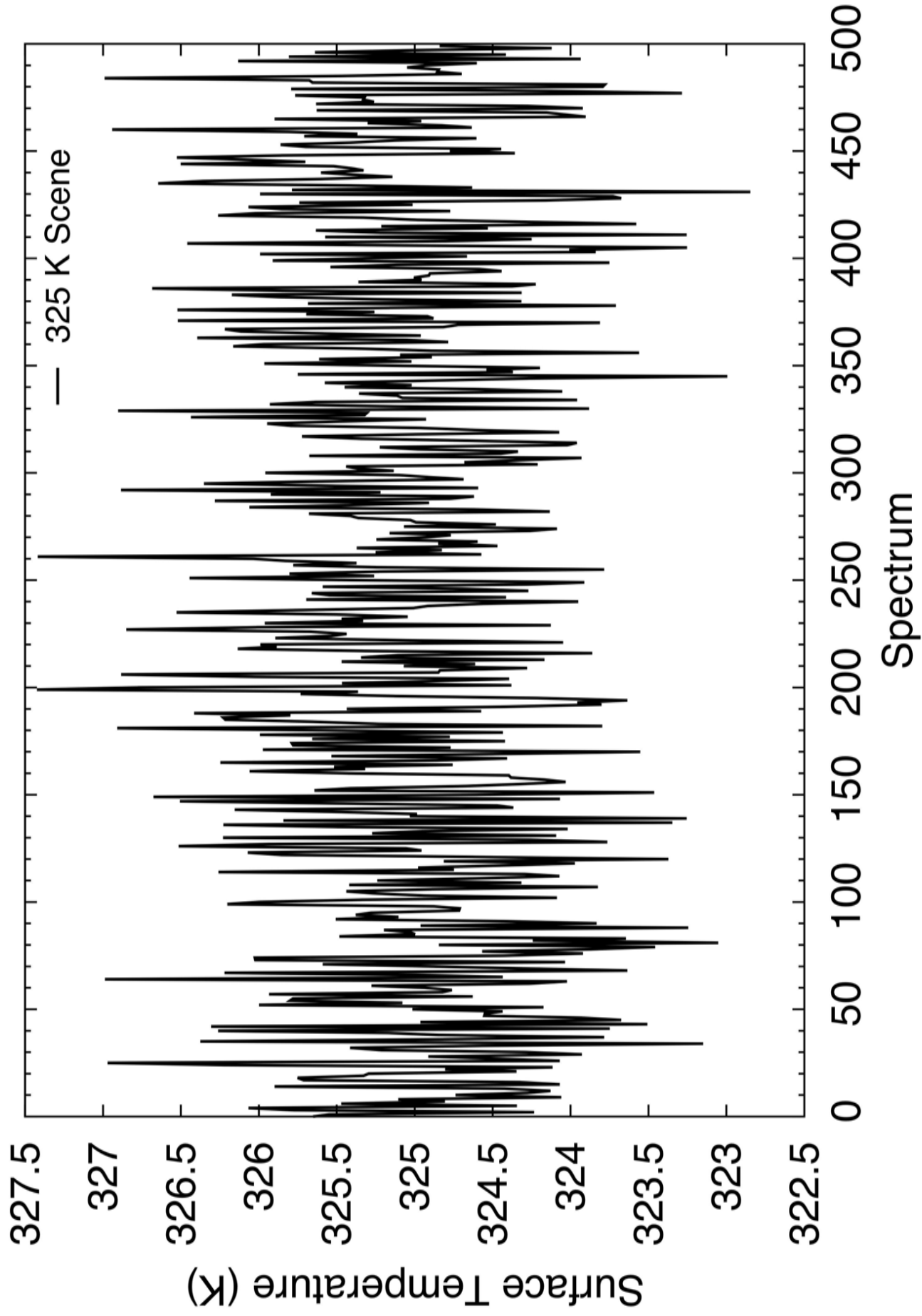

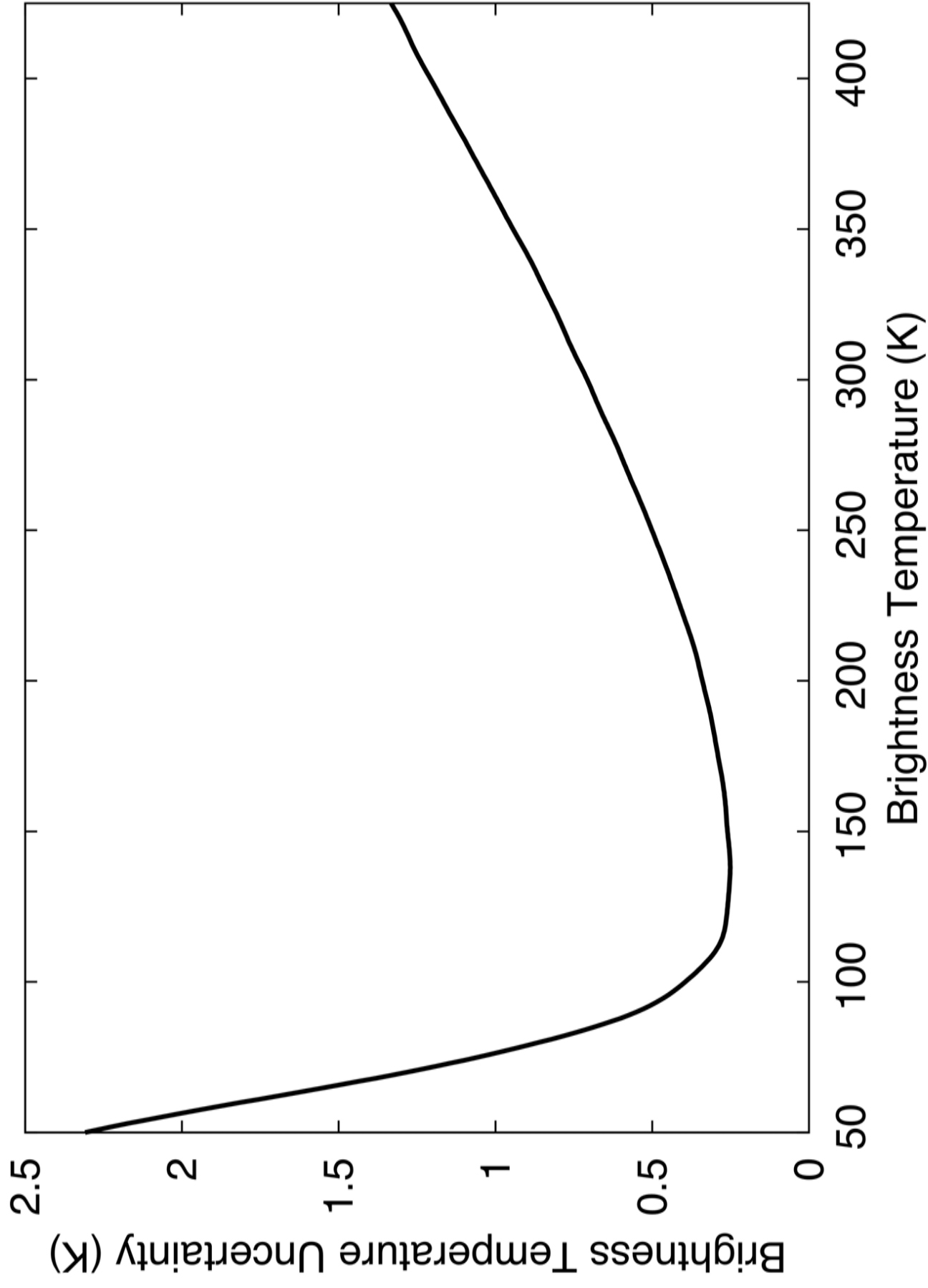

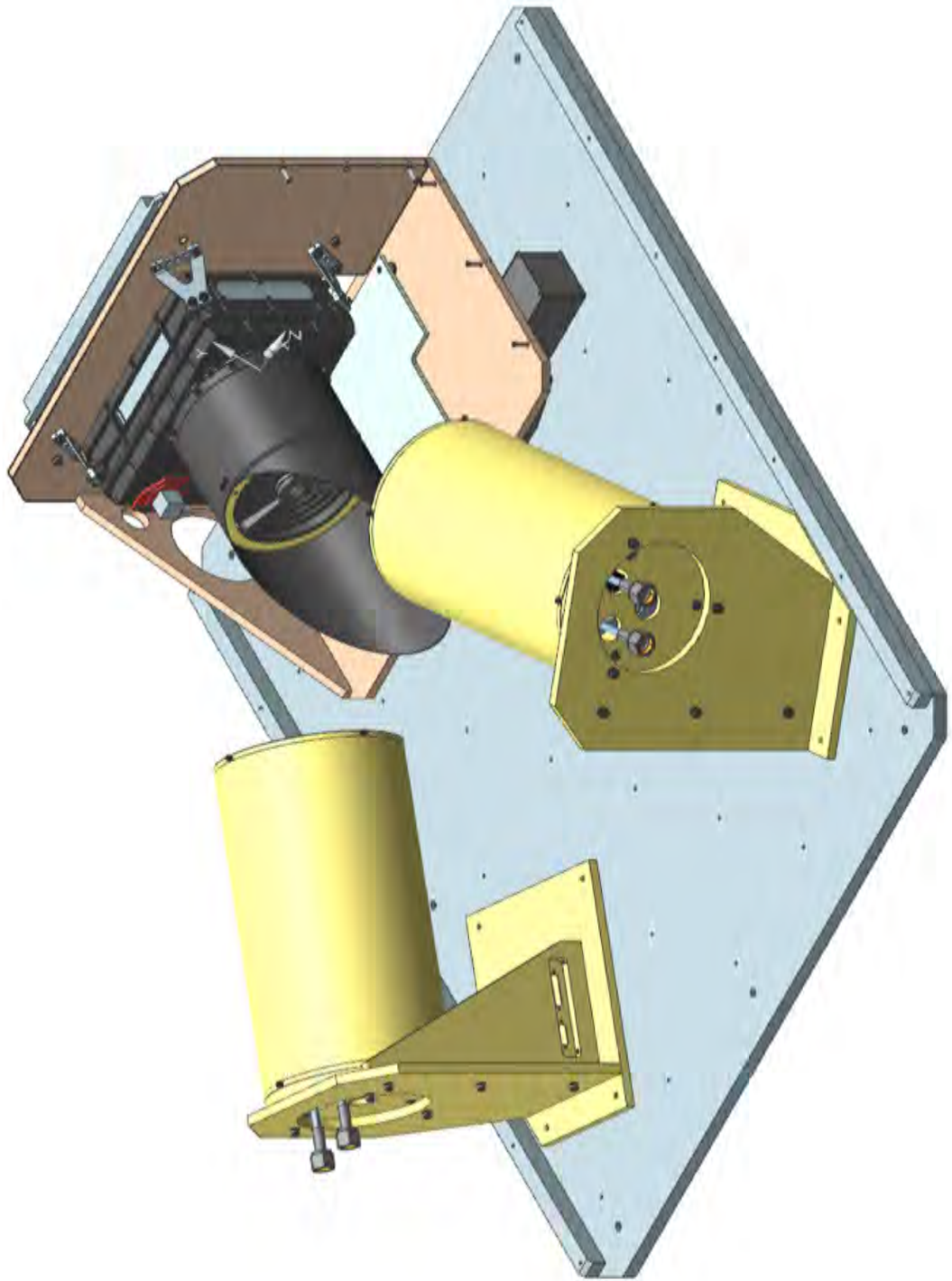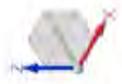

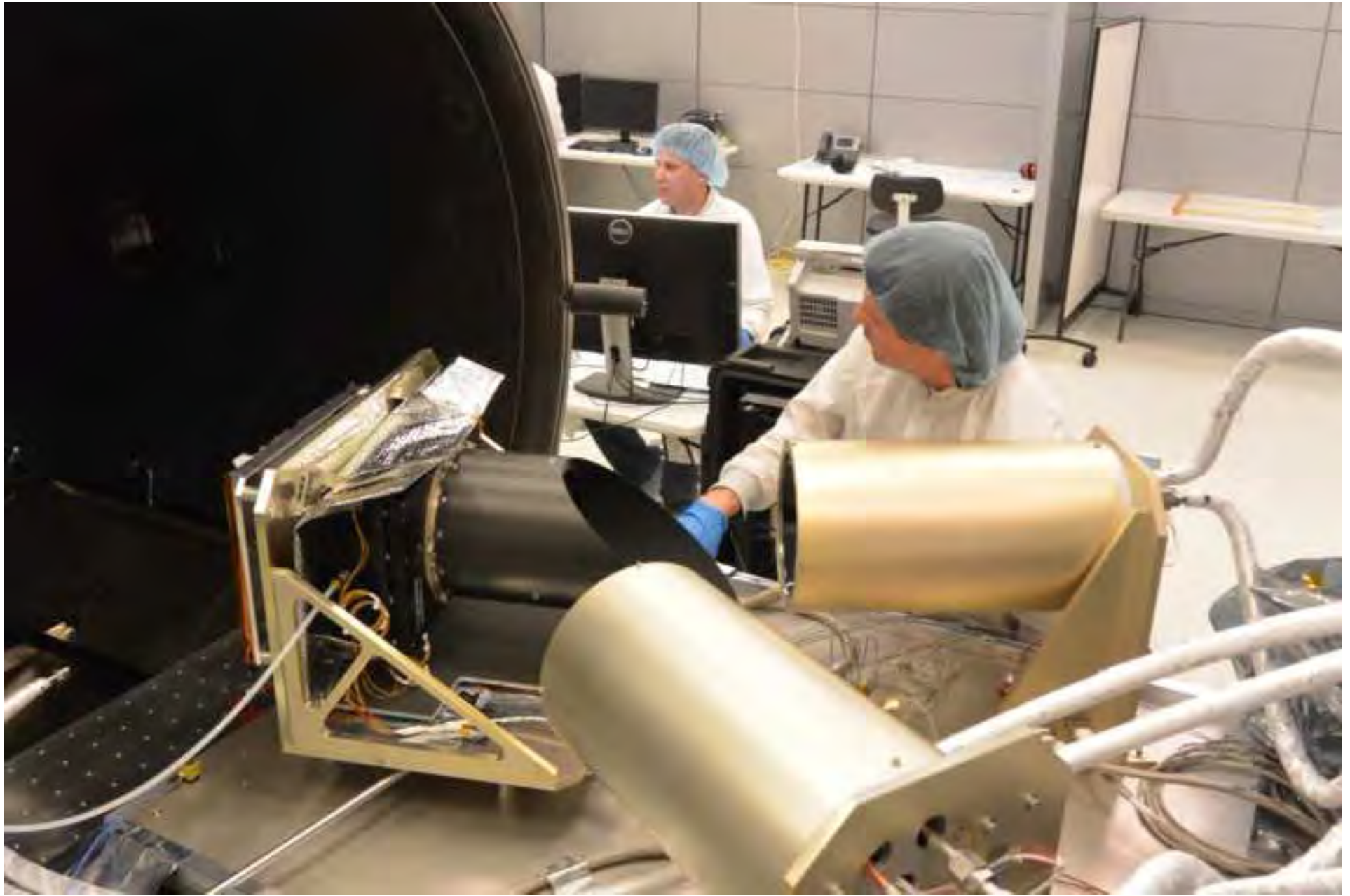

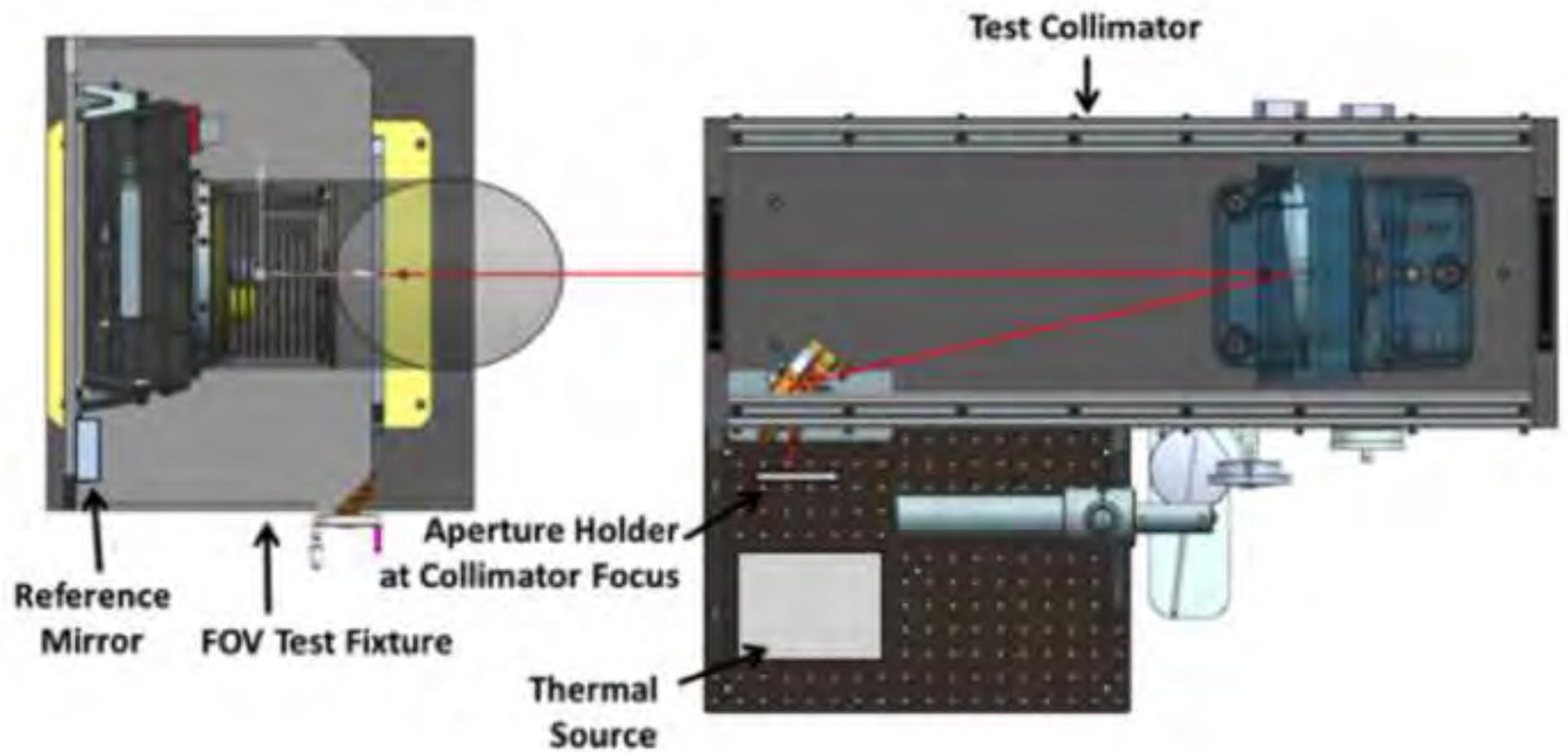

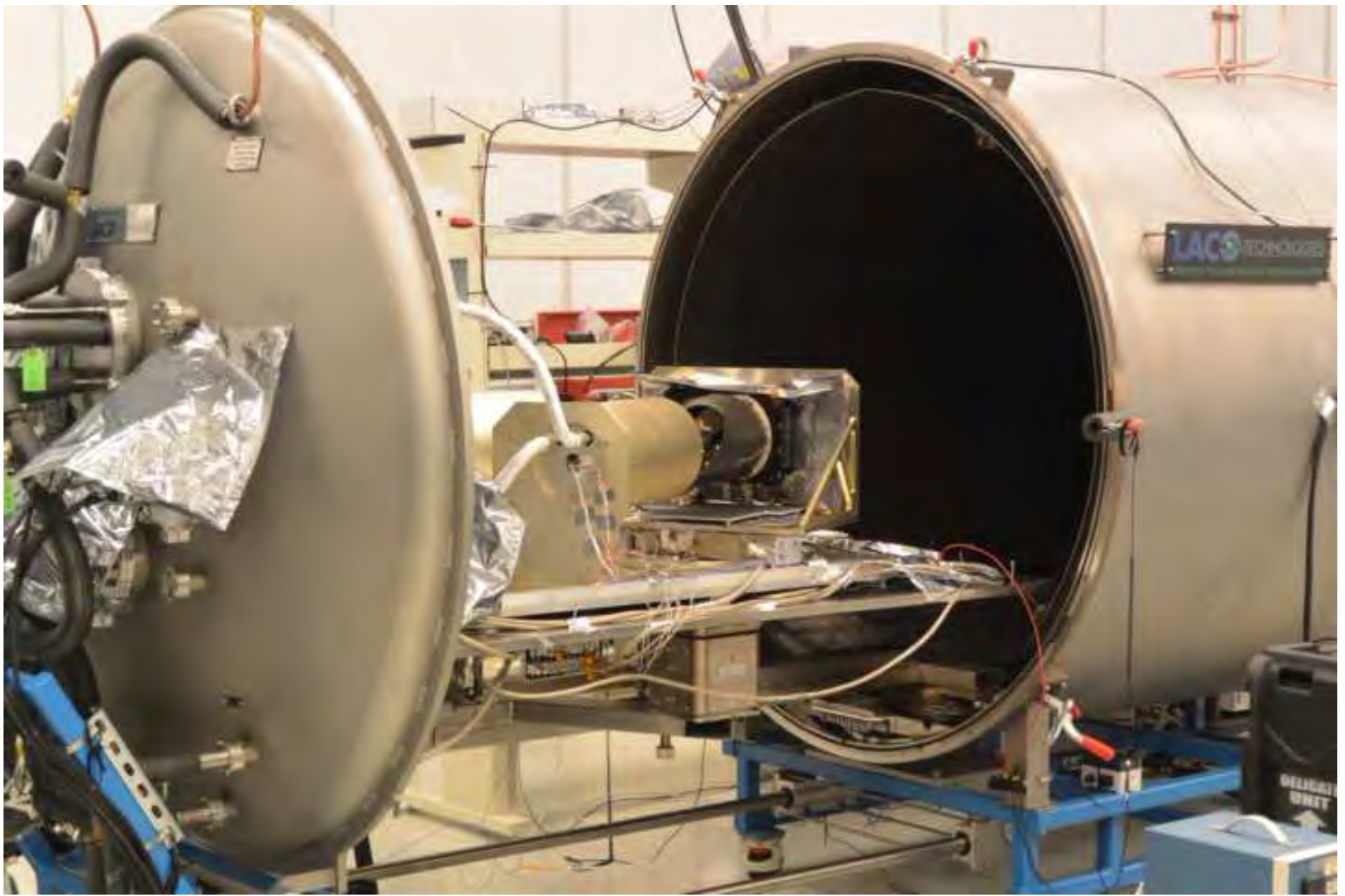

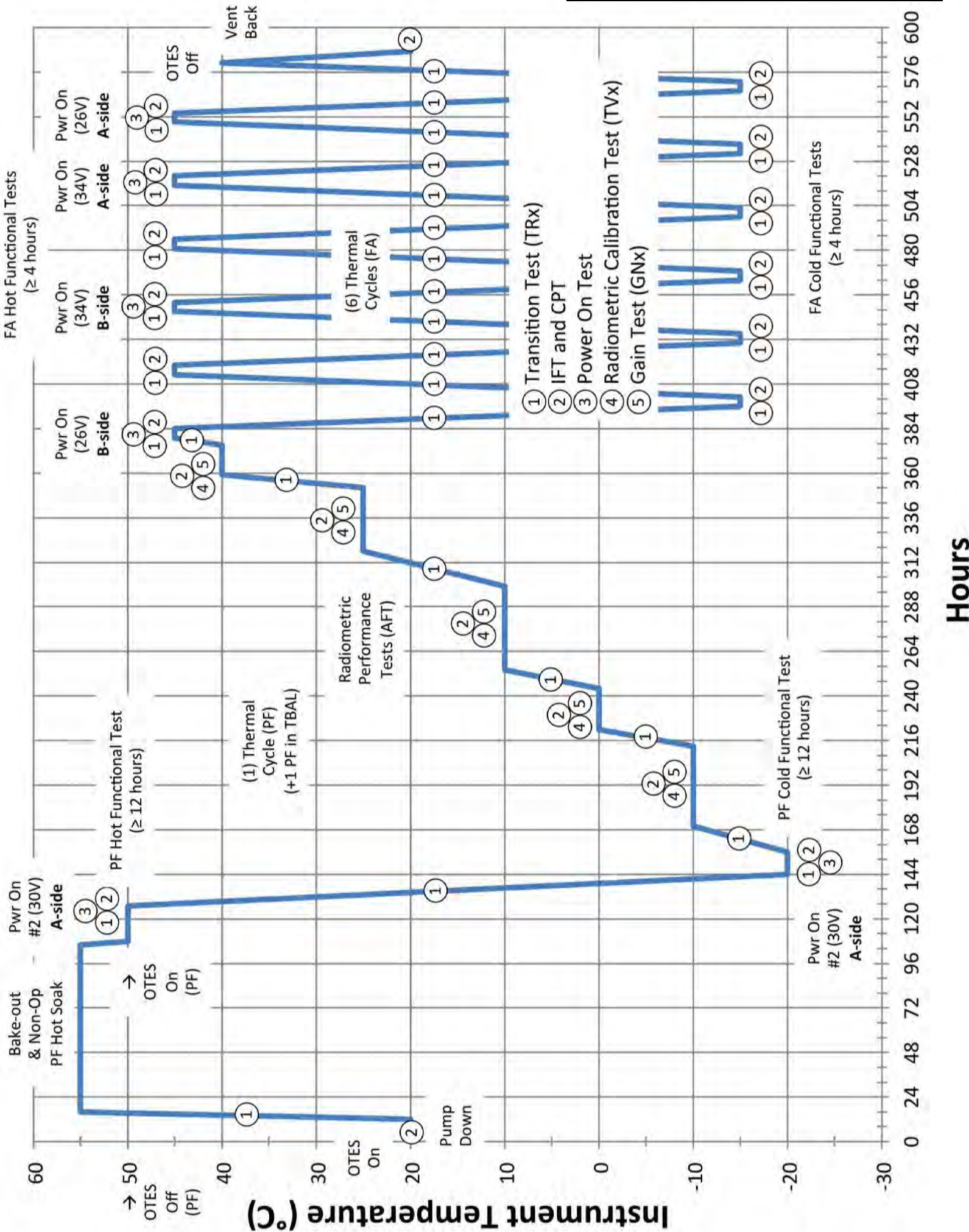

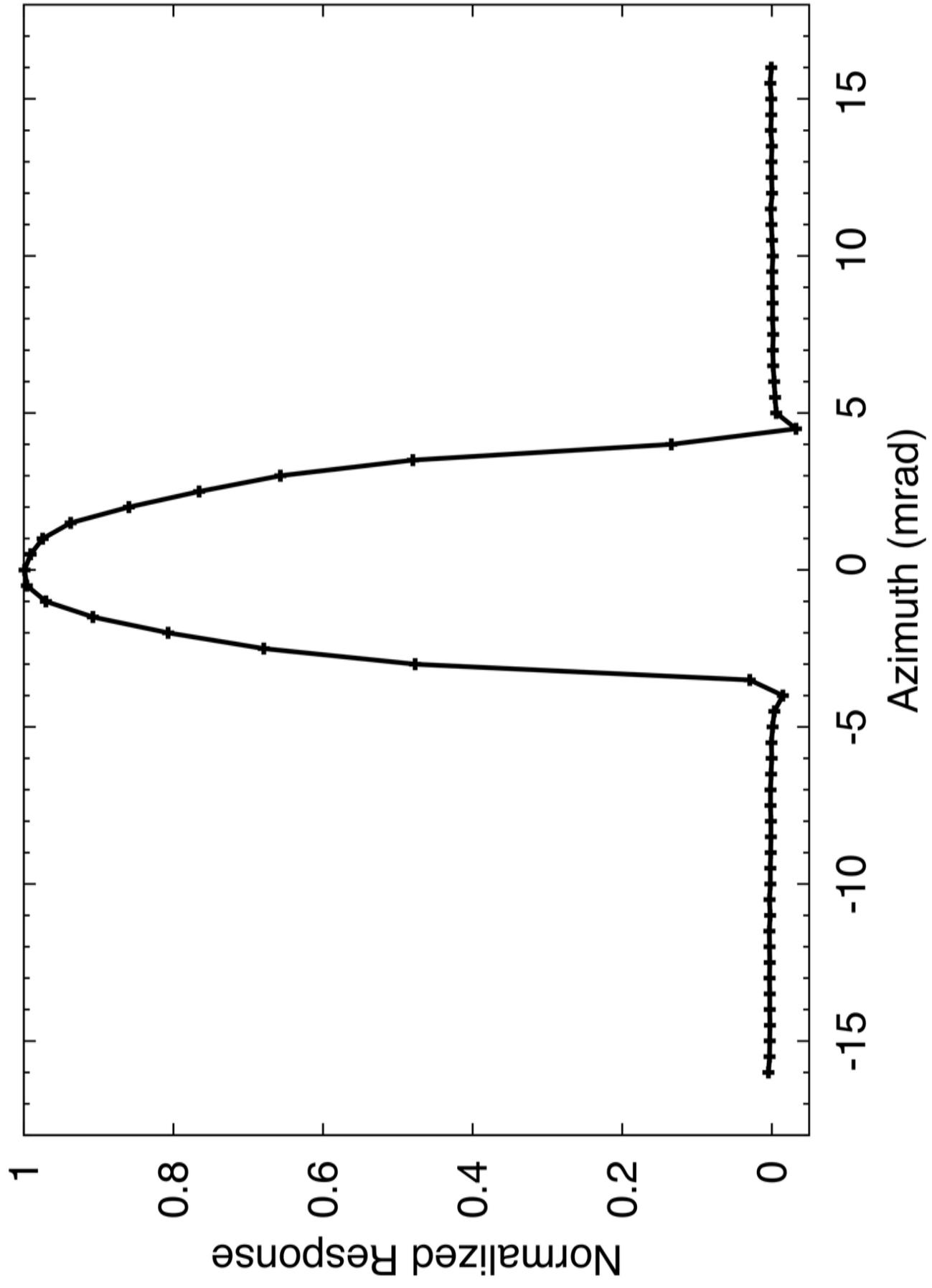

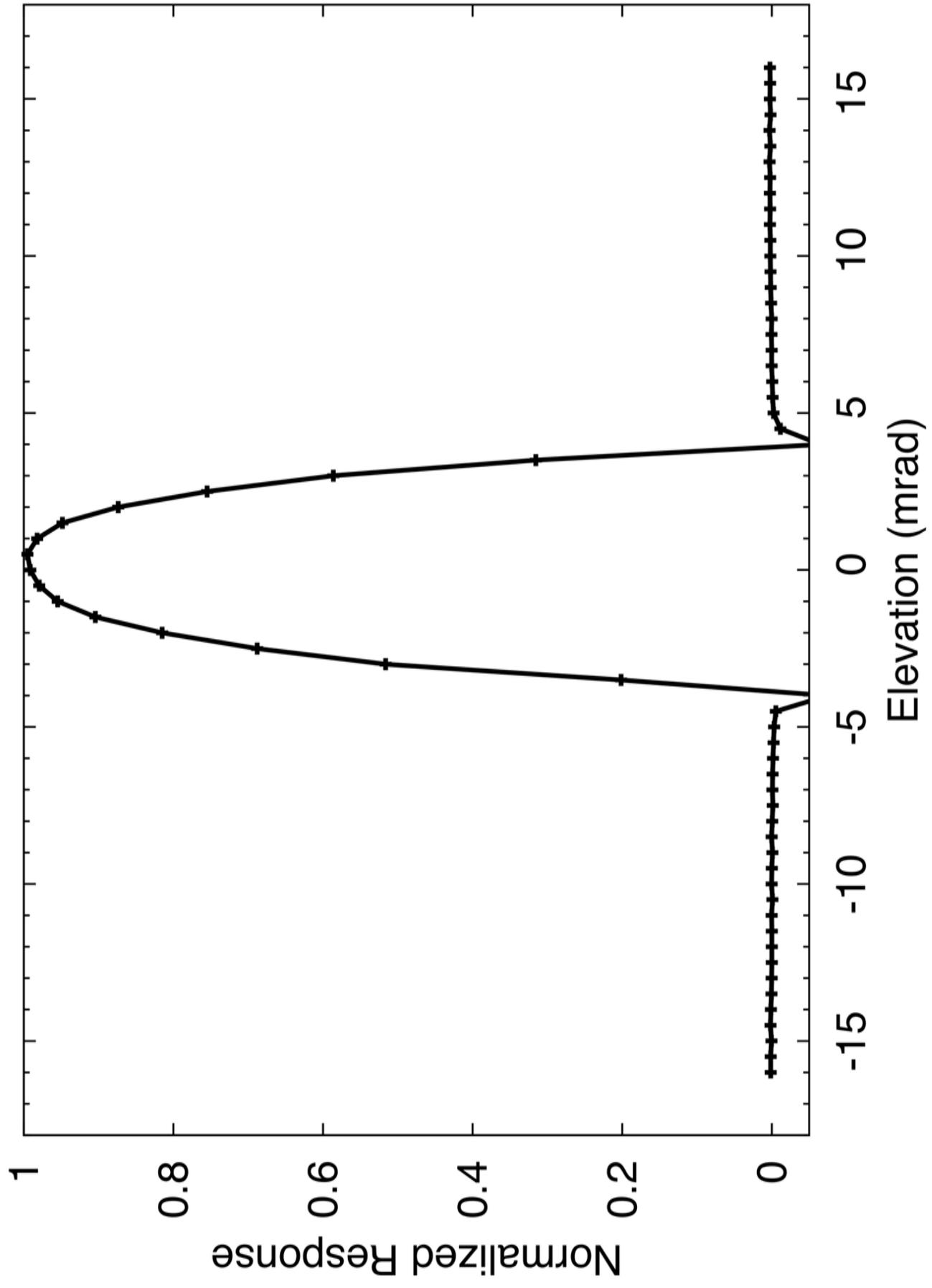

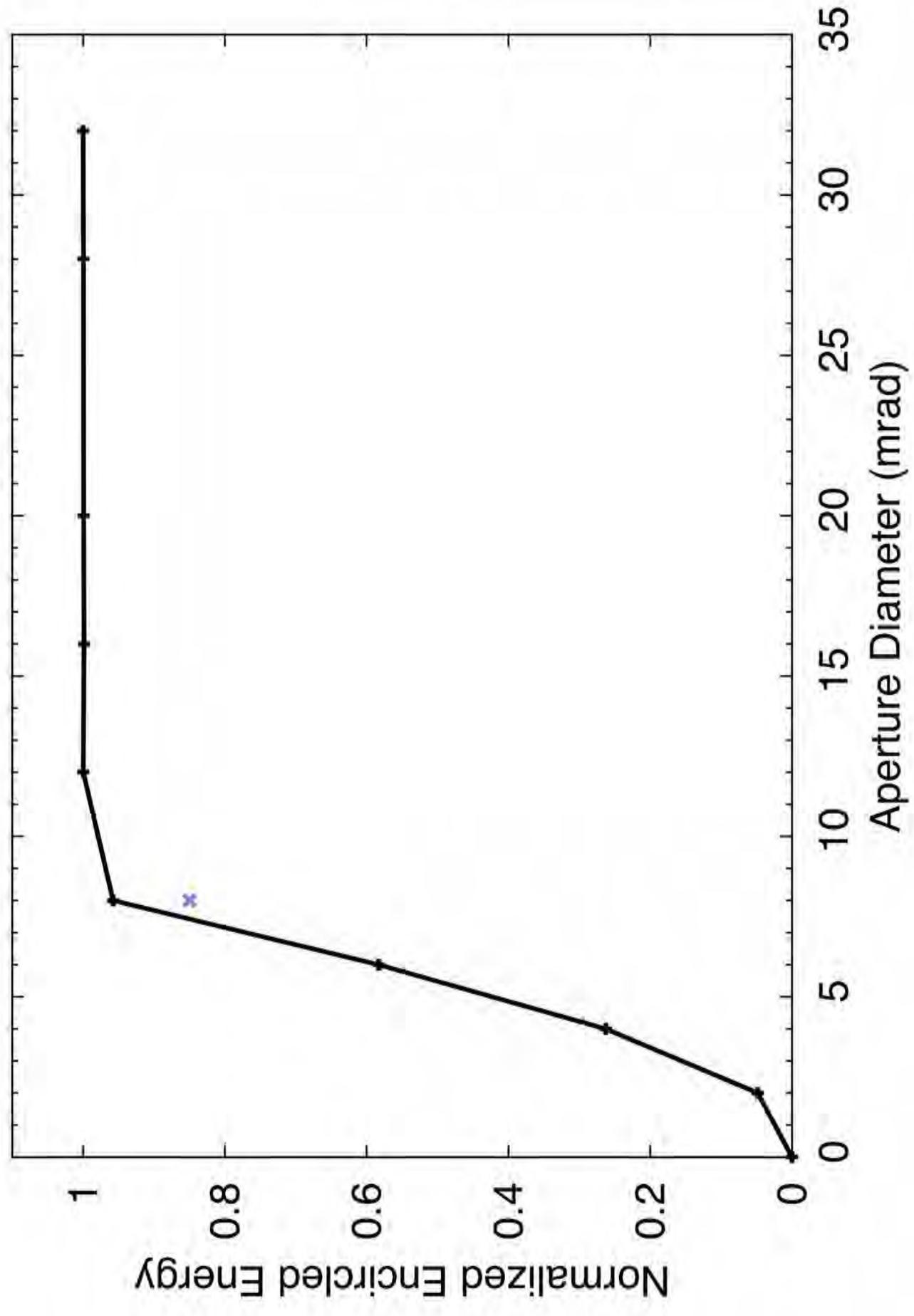

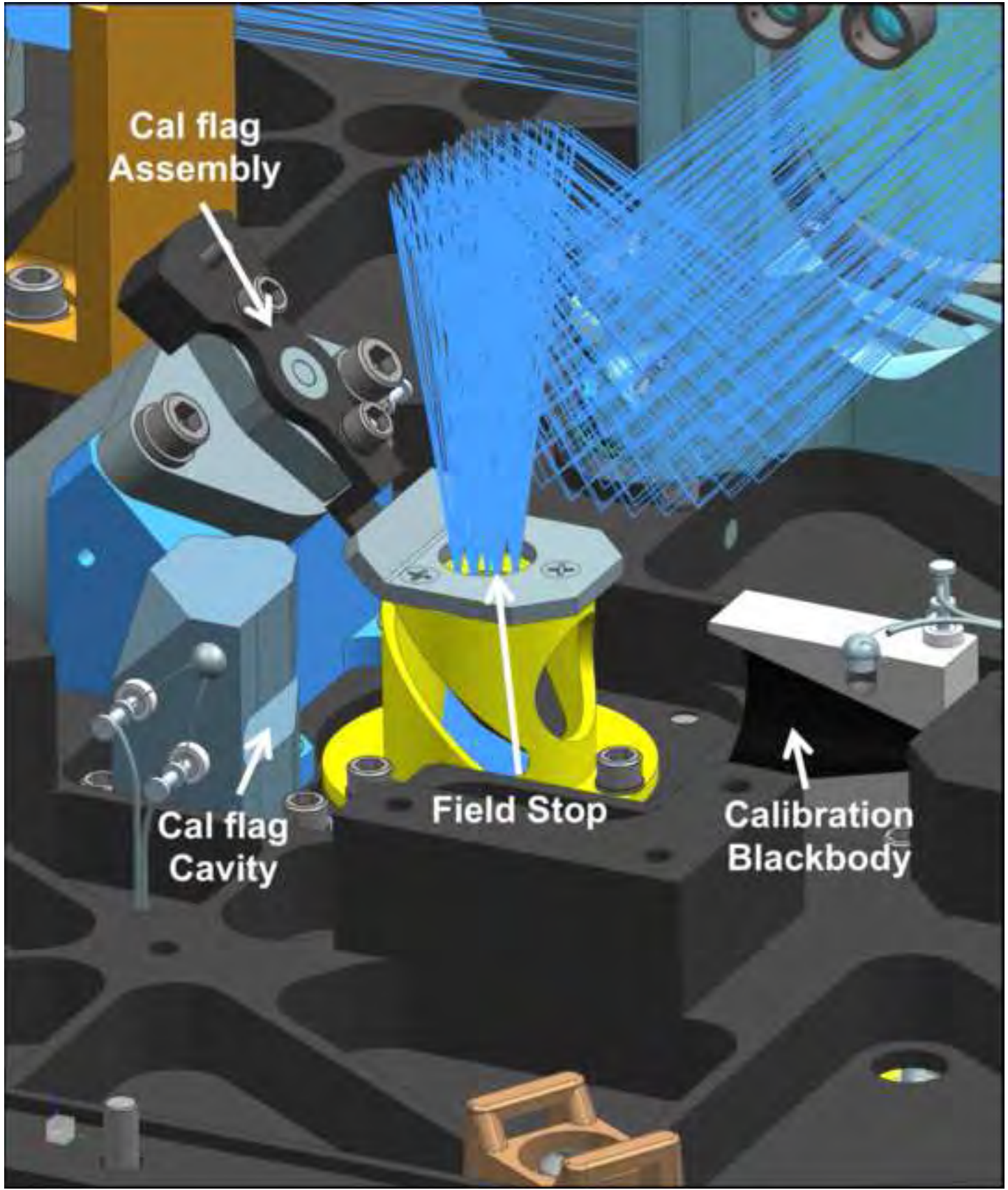

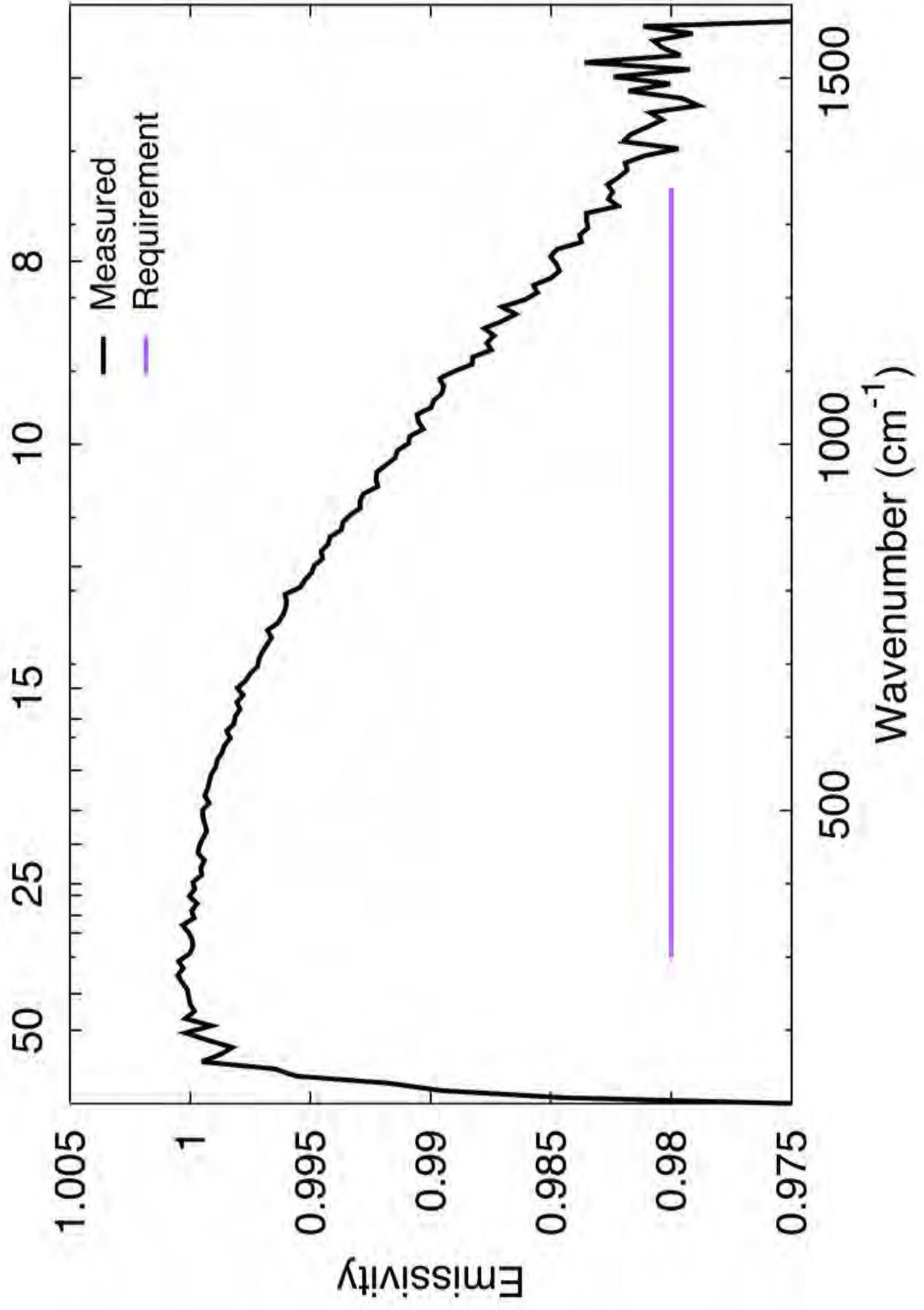

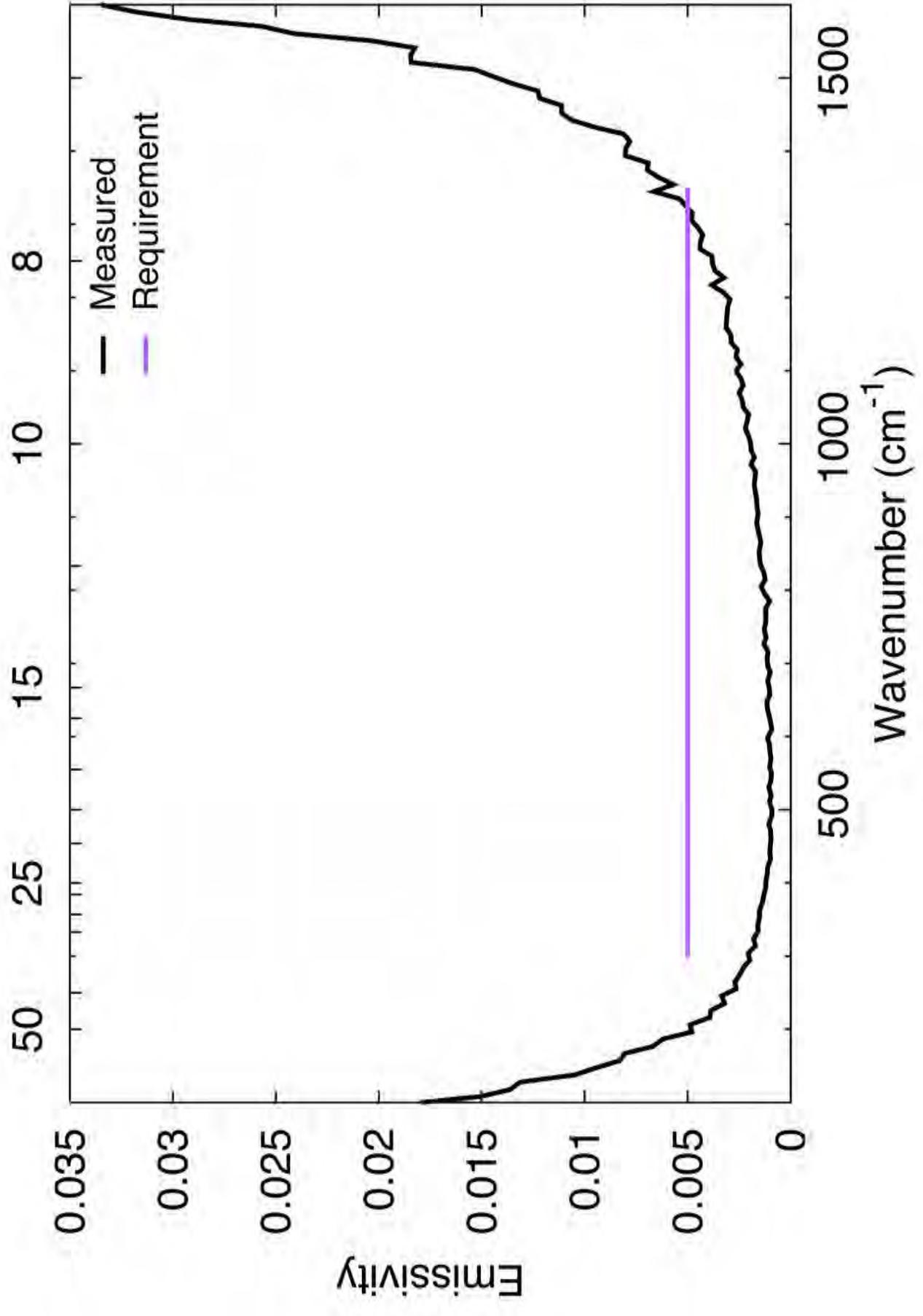

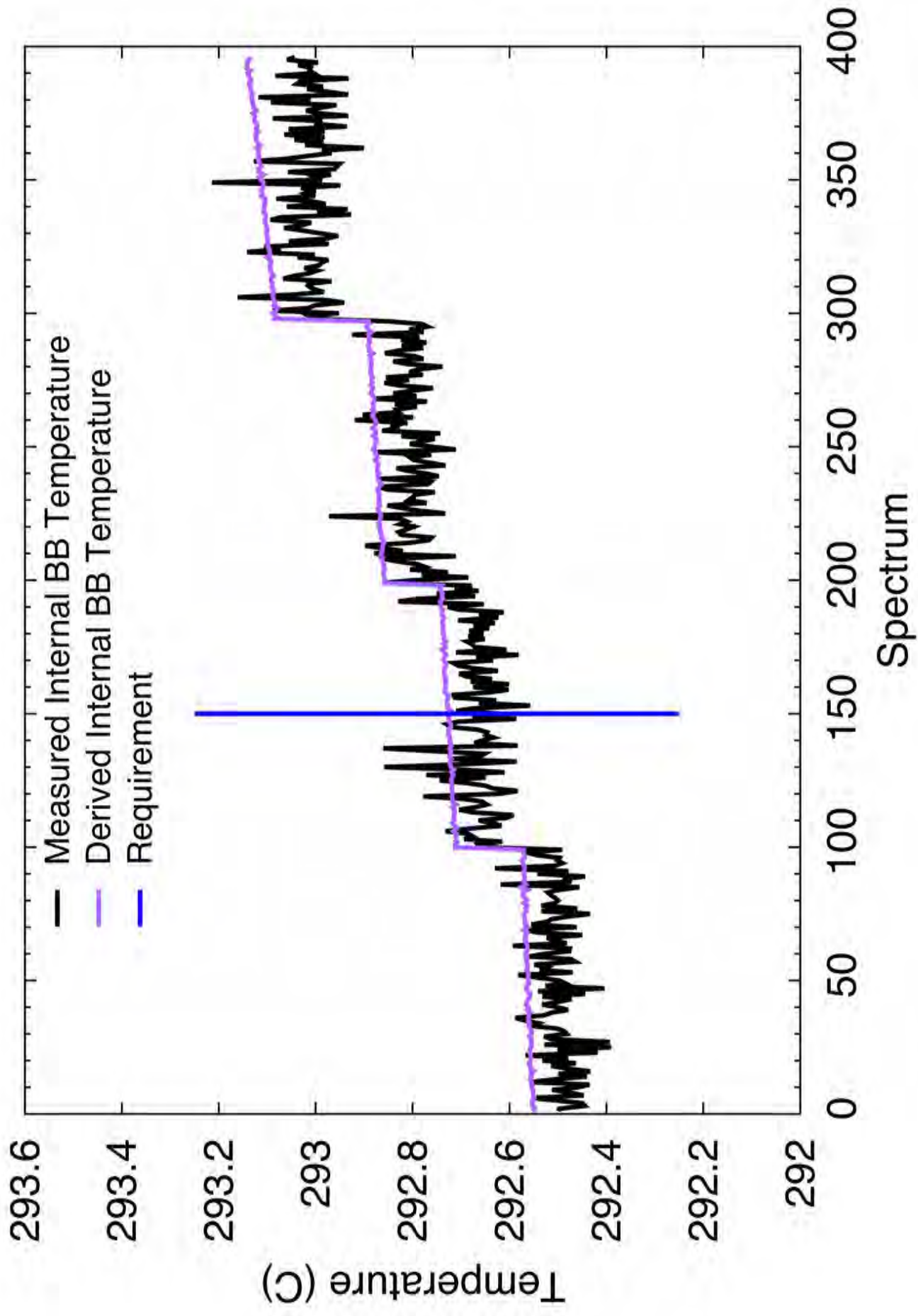

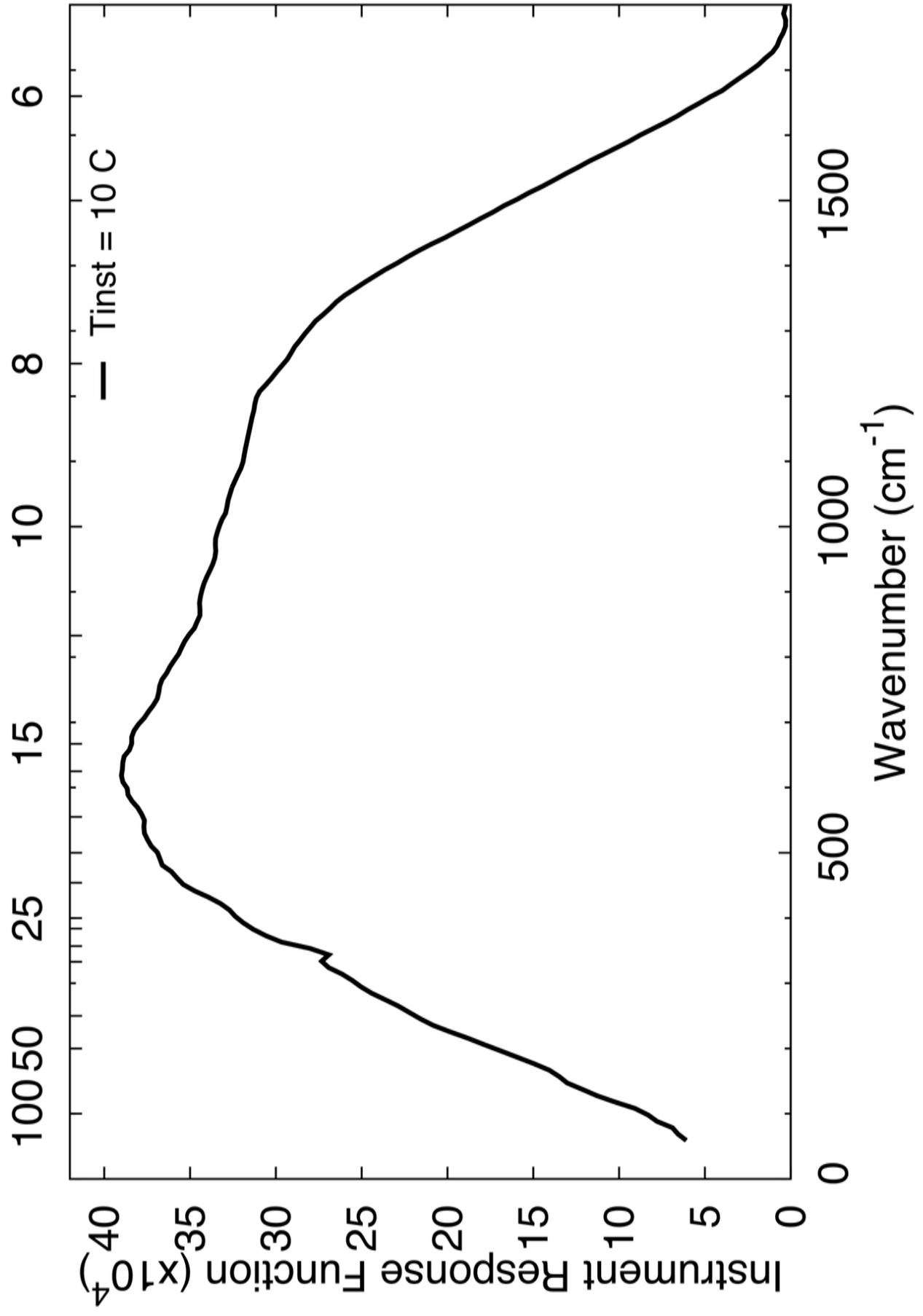

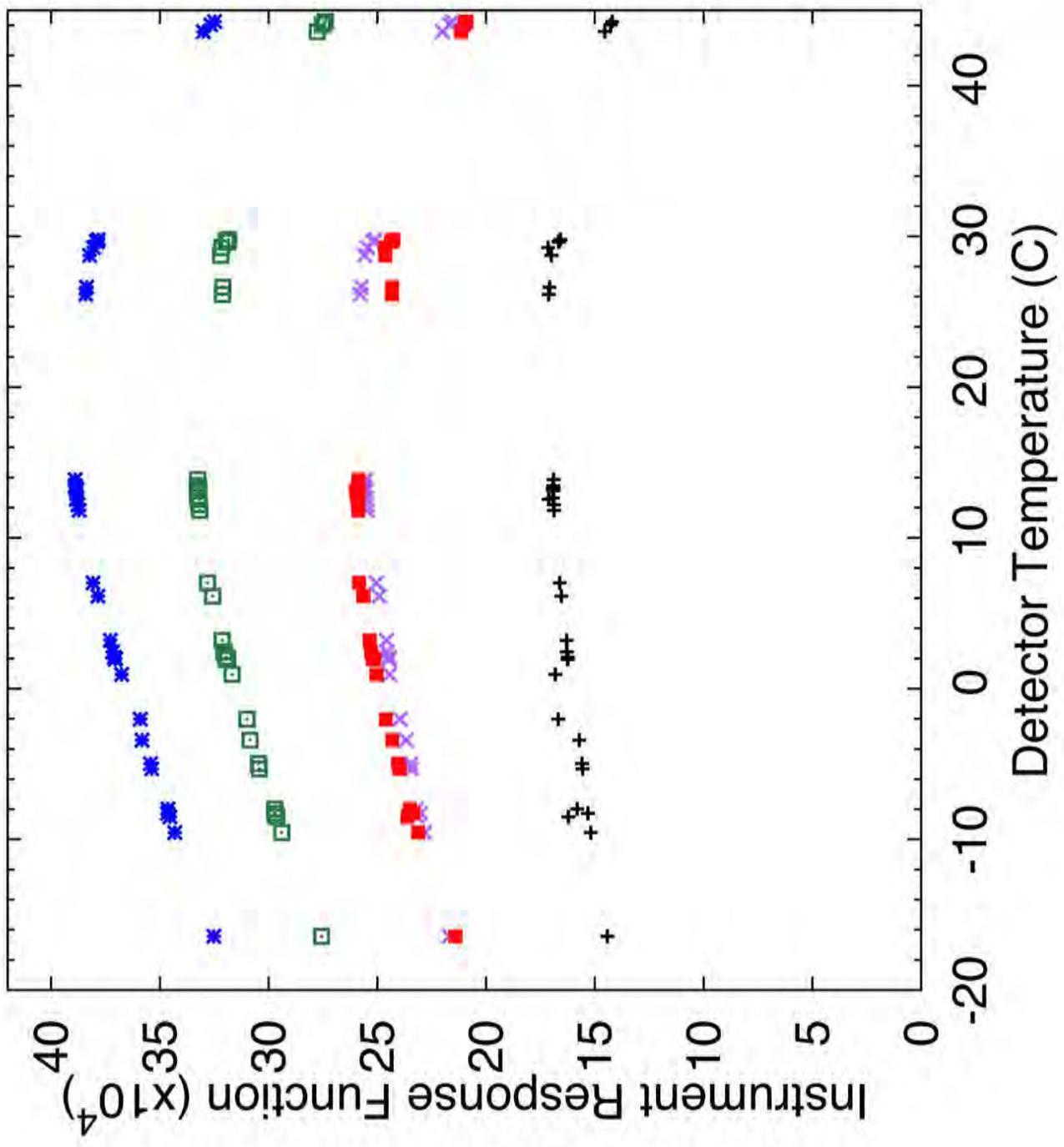

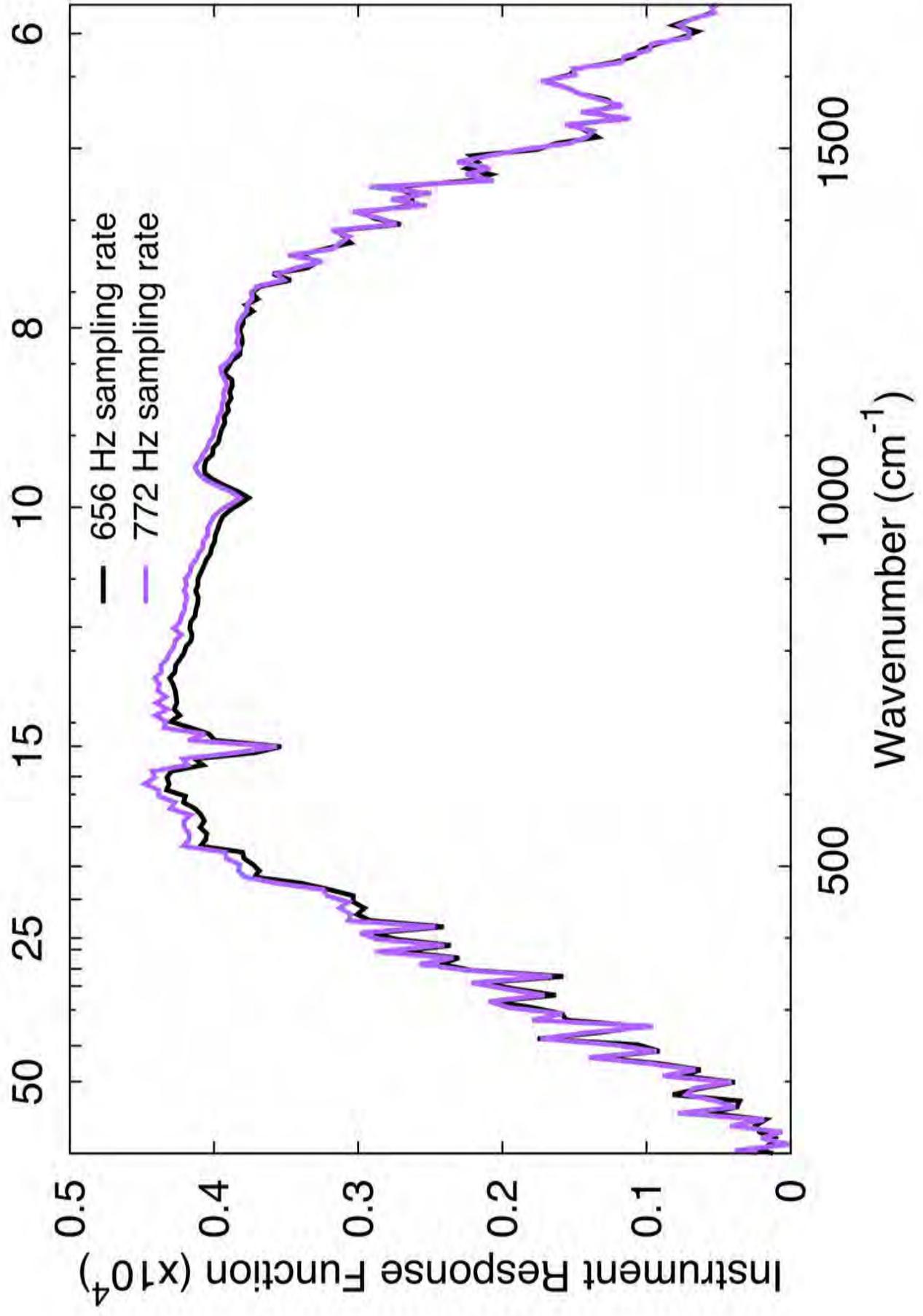

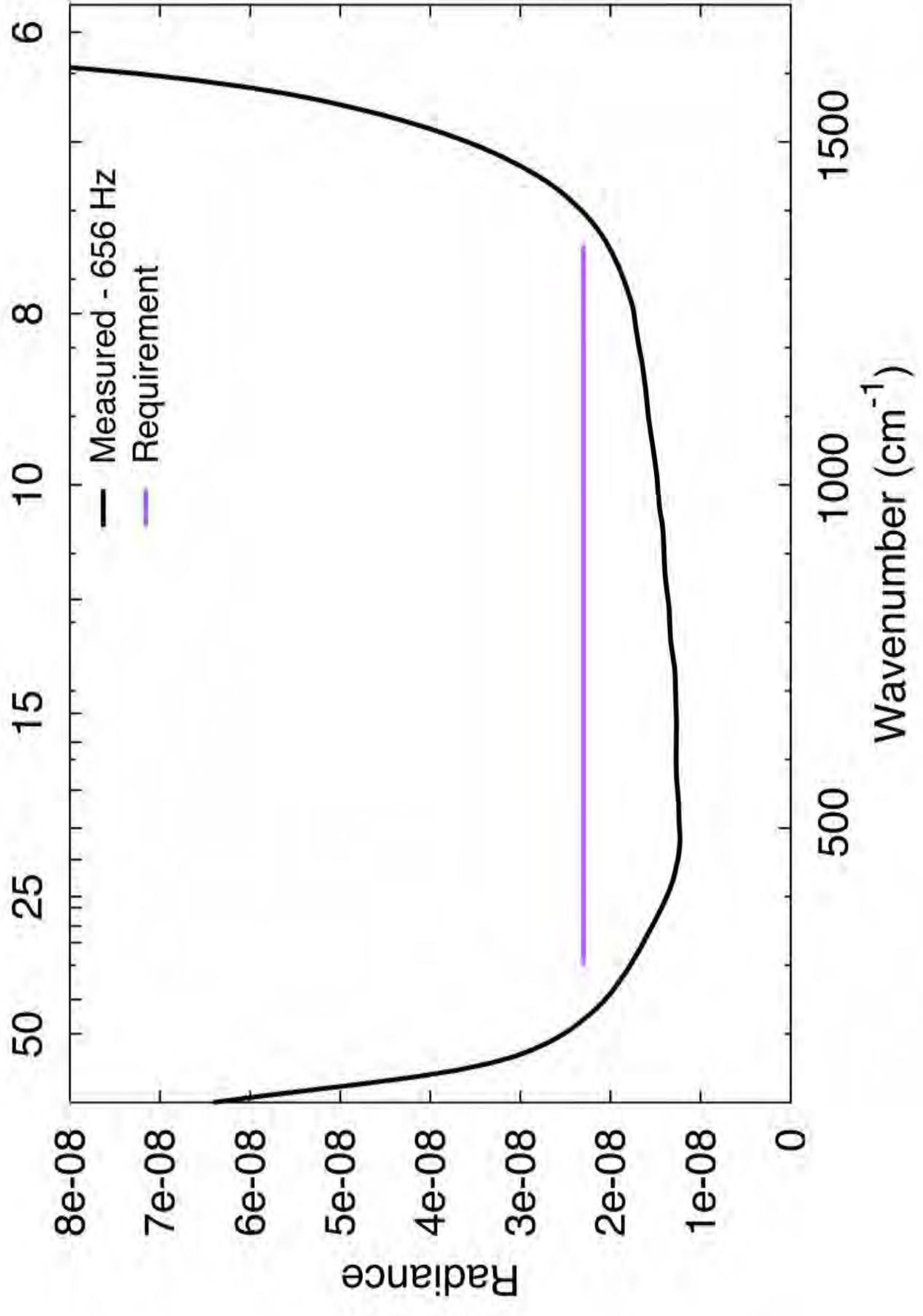

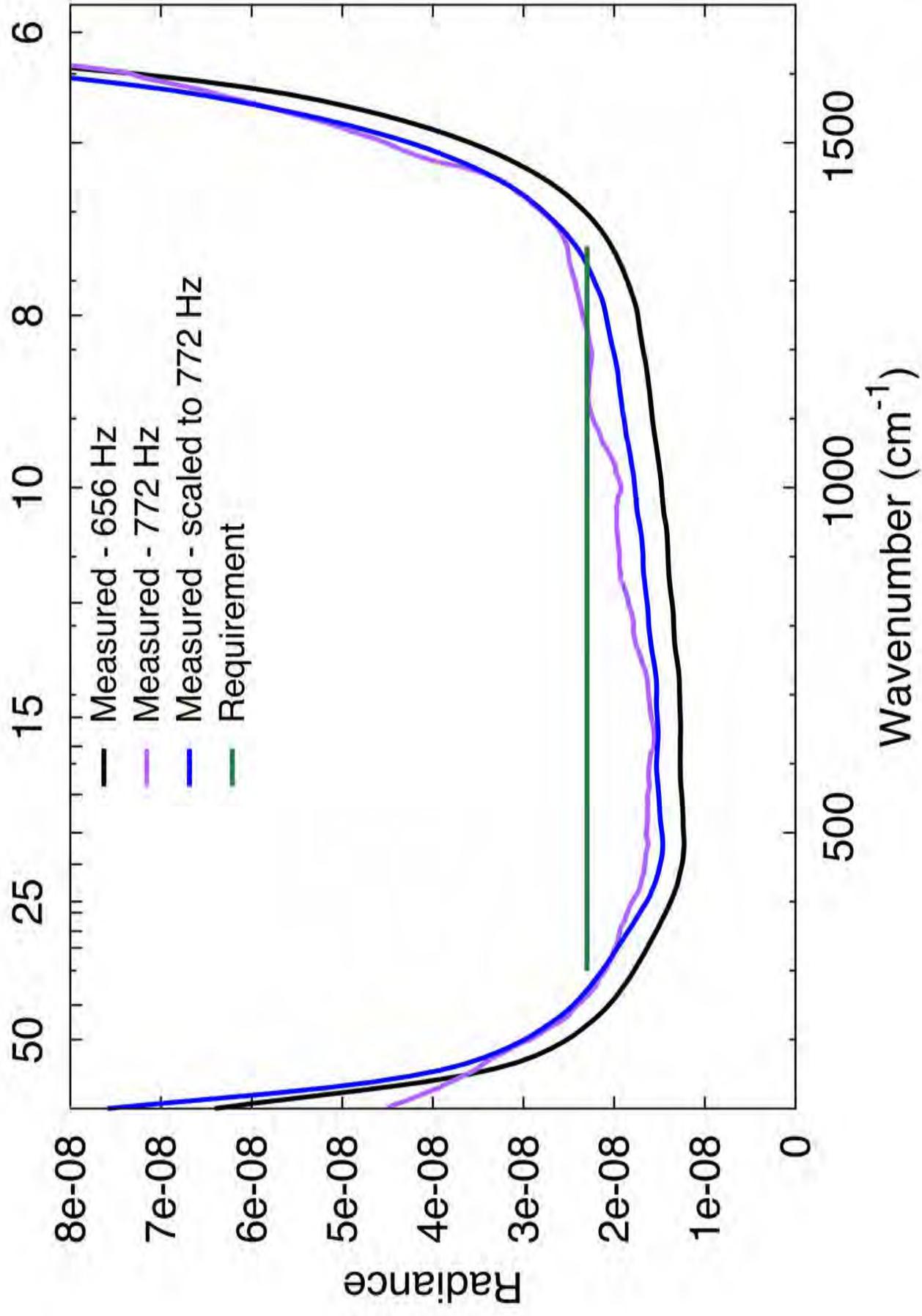

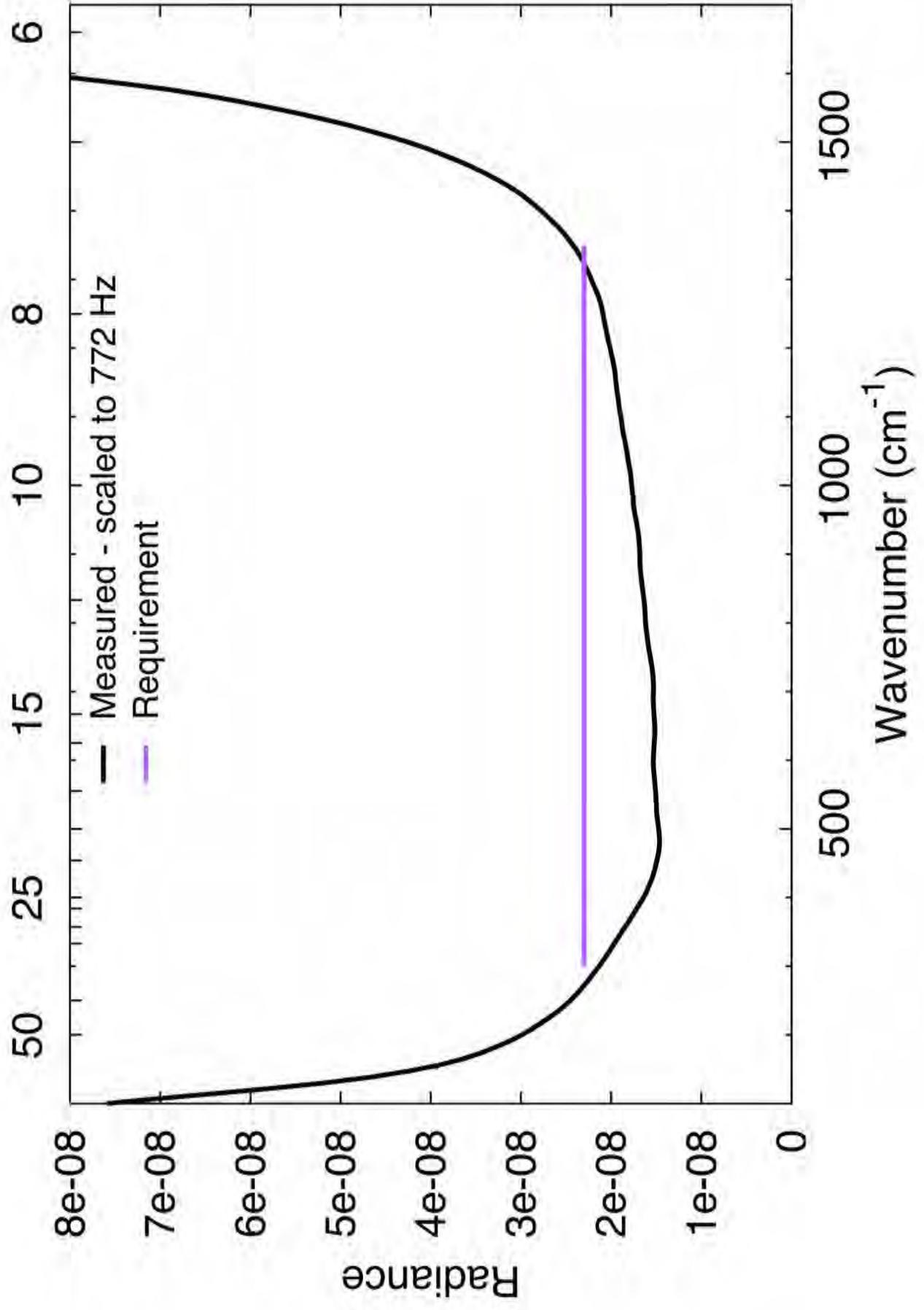

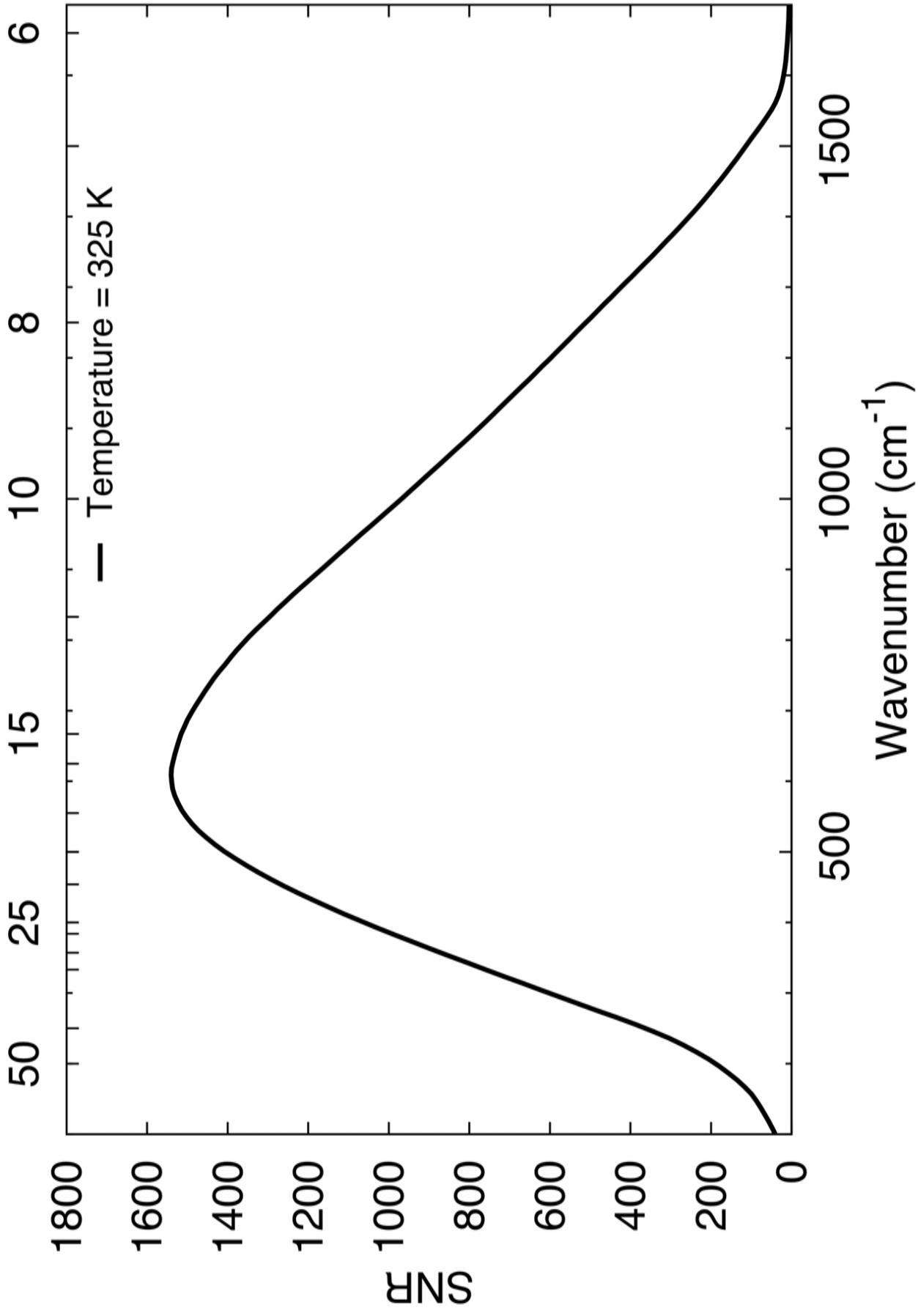

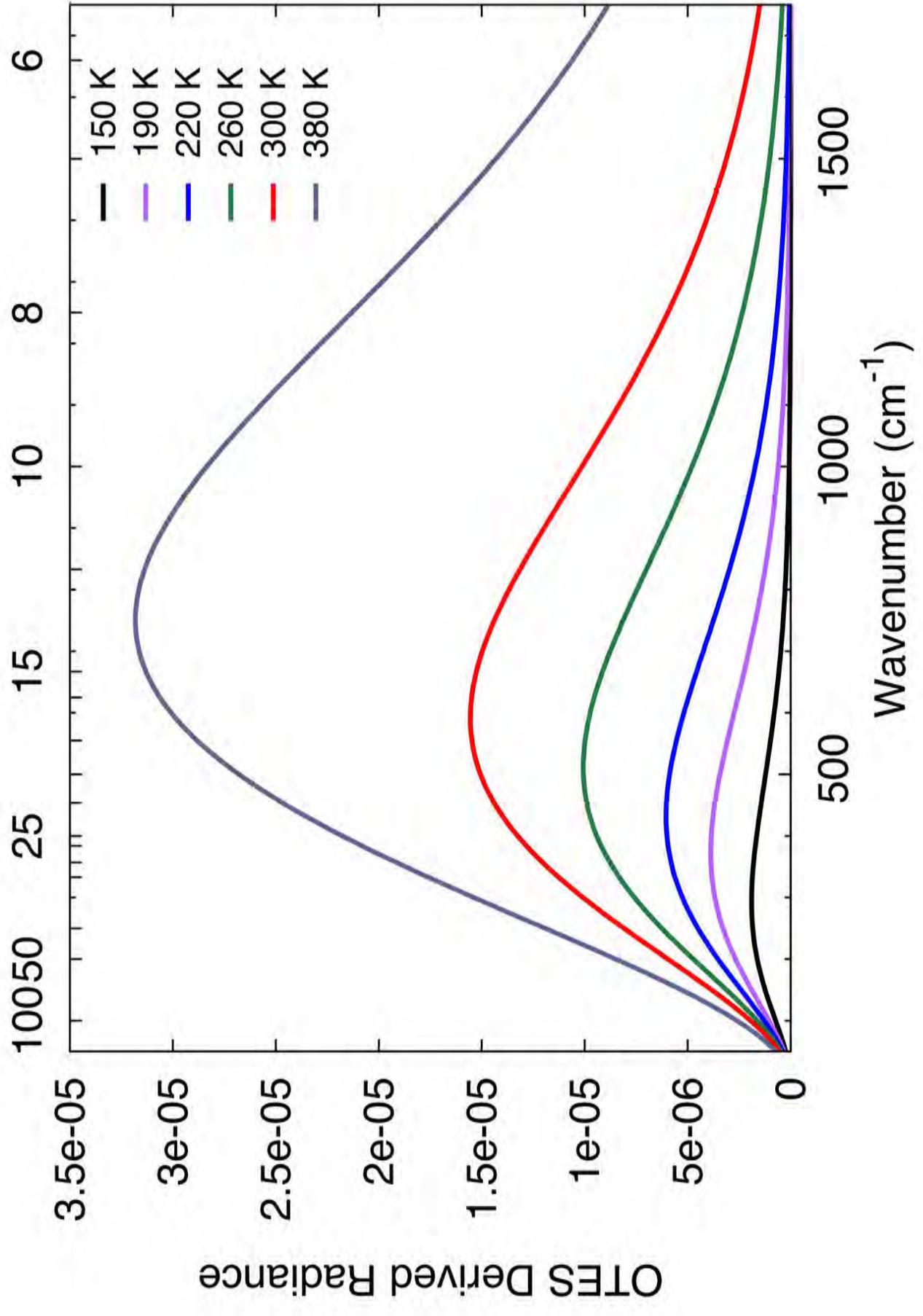

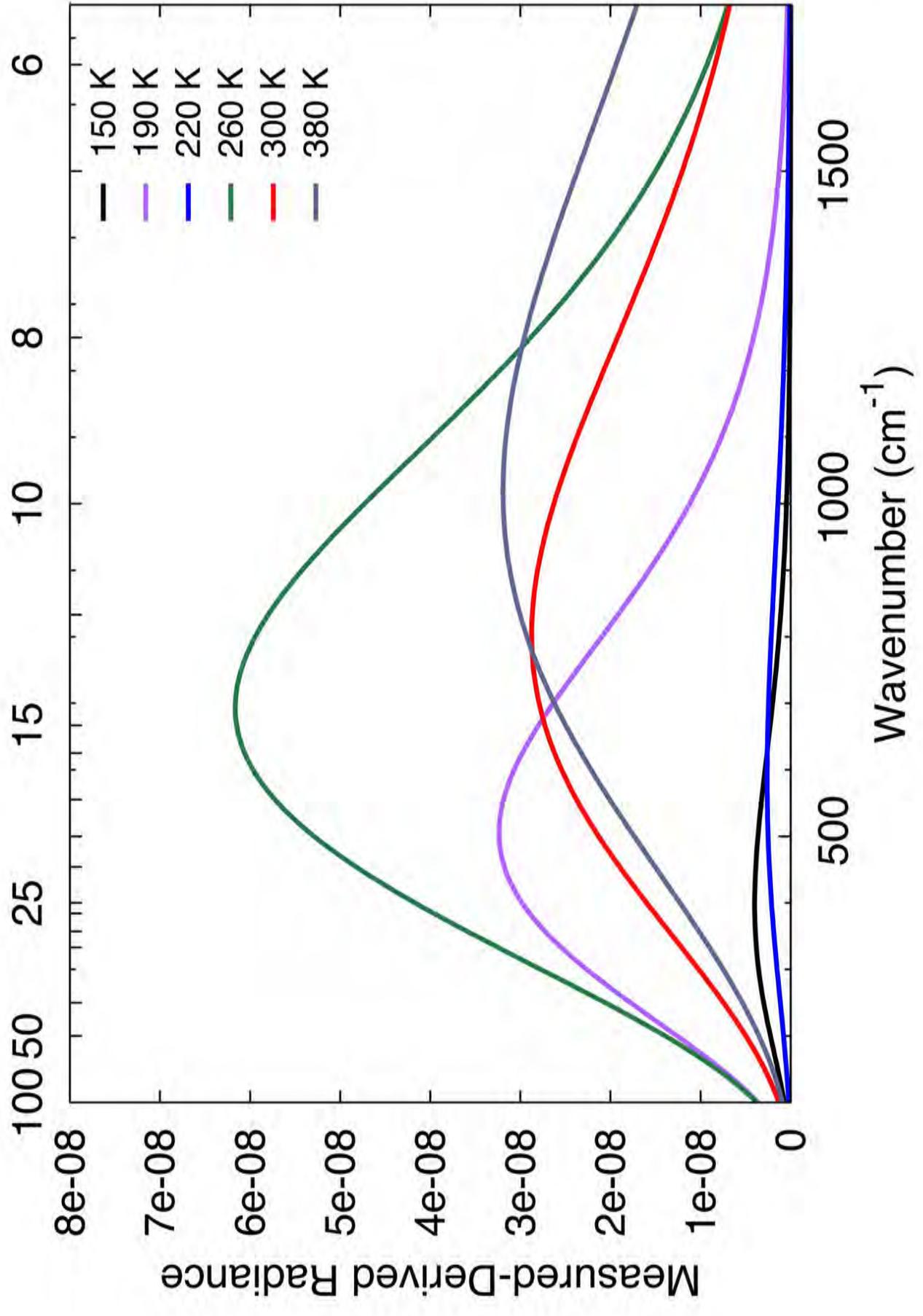

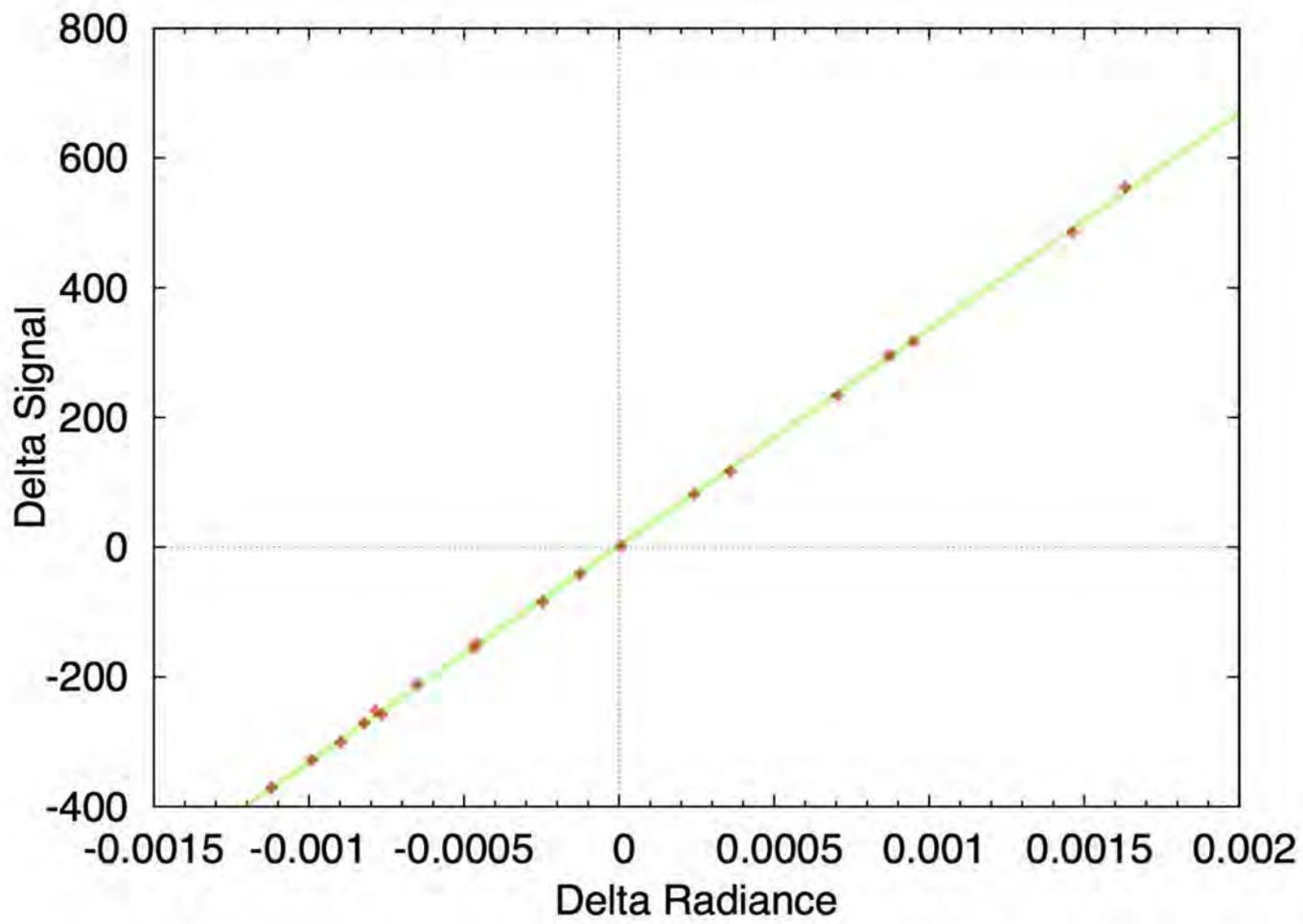

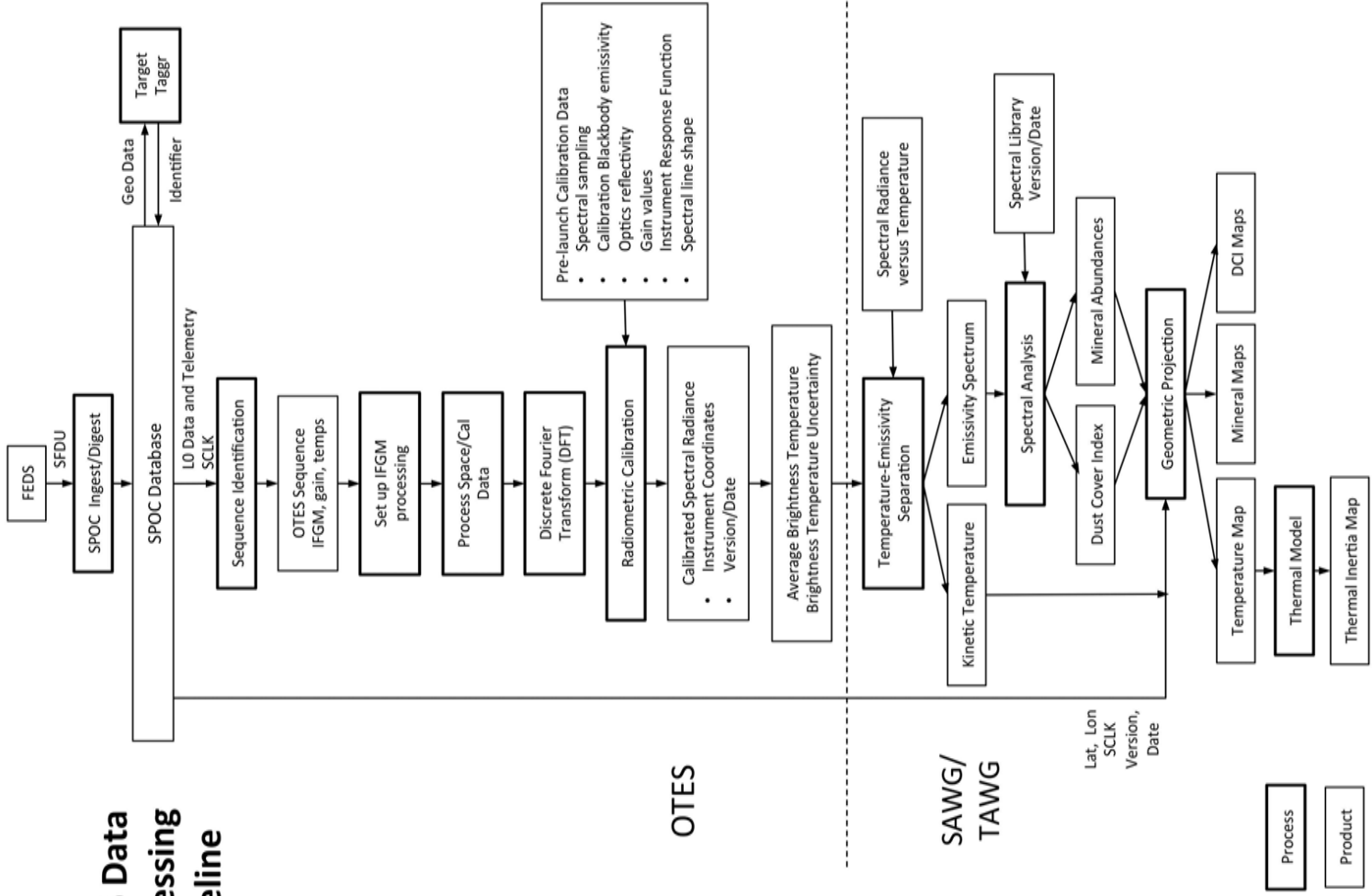